\documentclass[british,english]{article}
\usepackage[T1]{fontenc}
\usepackage[latin9]{inputenc}
\usepackage{geometry}
\usepackage{verbatim}
\usepackage{float}
\usepackage{mathrsfs}
\usepackage{amsmath}
\usepackage{amsthm}
\usepackage{amssymb}
\usepackage{graphicx}
\usepackage[numbers]{natbib}

\makeatletter

\providecommand{\tabularnewline}{\\}
\floatstyle{ruled}
\newfloat{algorithm}{tbp}{loa}
\providecommand{\algorithmname}{Algorithm}
\floatname{algorithm}{\protect\algorithmname}

\theoremstyle{plain}
\newtheorem{thm}{\protect\theoremname}
  \theoremstyle{remark}
  \newtheorem{rem}[thm]{\protect\remarkname}

\usepackage{babel}
\date{}

\makeatother

\usepackage{babel}
  \addto\captionsbritish{\renewcommand{\remarkname}{Remark}}
  \addto\captionsbritish{\renewcommand{\theoremname}{Theorem}}
  \addto\captionsenglish{\renewcommand{\remarkname}{Remark}}
  \addto\captionsenglish{\renewcommand{\theoremname}{Theorem}}
  \providecommand{\remarkname}{Remark}
\addto\captionsbritish{\renewcommand{\algorithmname}{Algorithm}}
\providecommand{\theoremname}{Theorem}

\begin{document}

\title{Particle Filtering for Stochastic Navier-Stokes Signal Observed
with Linear Additive Noise}

\author{Francesc Pons Llopis\thanks{Department of Mathematics, Imperial College London, UK.}\,\,\,, 
Nikolas Kantas$^{*}$, Alexandros Beskos\thanks{Department of Statistical Science, University College London, UK.},
Ajay Jasra\thanks{Department of Statistics and Applied Probability, National University
of Singapore}}
\maketitle
\begin{abstract}
We consider a non-linear filtering problem, whereby the signal obeys
the stochastic Navier-Stokes equations and is observed through a linear
mapping with additive noise. The setup is relevant to data assimilation
for numerical weather prediction and climate modelling, where similar models are used for unknown ocean or wind velocities.
We present a particle filtering methodology that uses likelihood informed
importance proposals, adaptive tempering, and a small number of appropriate
Markov Chain Monte Carlo steps. We provide a detailed design for each
of these steps and show in our numerical examples that they are all 
 crucial in terms of achieving good performance and efficiency.
\end{abstract}

\section{Introduction}

We  focus on a stochastic filtering problem where a space
and time varying hidden signal is observed at discrete times with
noise. The non-linear filtering problem consists of
computing the conditional probability law of the hidden
stochastic process (the so-called signal) given observations of it
 collected in a sequential manner. In particular, we
model the signal with a particular dissipative stochastic partial
differential equation (SPDE), which is the stochastic Navier-Stokes
Equation (NSE). This model, or a variant thereof, is often used in applications to model unknown quantities such as atmosphere or ocean velocity. In the spirit of data assimilation and uncertainty
quantification, we wish to extract information for the trajectory
of the hidden signal from noisy observations using a Bayesian approach.
Typical applications include numerical weather forecasting in meteorology,
oceanography and atmospheric sciences,
geophysics, hydrology and petroleum engineering; see \citep{bennett2005inverse,majda2012filtering,bocquet2010beyond}
for an overview.

We restrict to the setting where the state of interest is the time
varying velocity field, $V(x,t)$, in some 2D bounded
set $\Omega$. The unknown state is modelled using the stochastic
NSE 
\begin{equation}
dV(x,t)-\nu\Delta V(x,t)dt+B(V,V)(x,t)dt=f(x,t)dt+Q^{\frac{1}{2}}dW(x,t),\label{eq:spde-dynamics}
\end{equation}
where $\Delta$ is the Laplacian, $\nu$ a viscosity constant,
$B$ a non-linear operator due to convection, $Q$ a positive,
self adjoint, trace class operator, $f$ a determistic forcing and
$W(x,t)$ a space-time white noise as in \citep{da2008stochastic}.
This might appear as a restrictive choice for the dynamics, but the
subsequent methodology is generic and could be potentially applied
to other similar dissipative SPDEs, such as the stochastic Burger's
or Kuramoto\textendash Sivashinski equations \citep{jardak2010comparison,chorin2004dimensional}.

The evolution of the unknown state of the SPDE is observed at discrete
times and generates a sequence of noisy observations $\mathcal{Y}_{n}=\left(Y_{t_{1}},\ldots Y_{t_{n}}\right)$.
In order to perform accurate estimation and uncertainty quantification,
we are  interested not just in approximating a single trajectory
estimate of the hidden state, but in the complete filtering distribution, 
\begin{equation}
\pi_{n}(\stackrel{\scriptscriptstyle\bullet}{{}})=\mathbb{P}\left[\left.V(\cdot,t_{n})\in{\stackrel{\scriptscriptstyle\bullet}{{}}}\right|\mathcal{Y}_{n}\right],\label{eq:filter_intro}
\end{equation}
that is, the conditional distribution of the state given all the observations
obtained up to current time $t_n$. The main objective is to compute the filtering
distribution as it evolves with time, which is an instance of the
stochastic filtering problem \citep{bain2008fundamentals}. The solution
of the problem can be formulated rigorously as a recursive Bayesian
inference problem posed on an appropriate function space \citep{law2015data}.
In contrast to standard filtering problems, the problem setup here
is particularly challenging: the prior consists of a complicated probability
law generated by the SPDE \citep{da2008stochastic} and observation
likelihoods on the high dimensional space of the signal tend to be very informative.

The aim of this paper is to propose Sequential Monte Carlo (SMC) methods
(also known as Particle Filters (PF)) that can approximate effectively
these conditional distributions. Computing the evolution of the filtering
distribution $\pi_{n}$ is not analytically tractable, except in linear
Gaussian settings. SMC is a generic Monte Carlo method that approximates
the sequence of $\pi_{n}$-s and their normalising constant $\mathbb{P}\left[\mathcal{Y}_{n}\right]$
(known in Statistics as marginal likelihood or evidence). This is achieved
by obtaining samples known as particles and combining Importance Sampling
(IS), resampling and parallel Markov Chain Monte Carlo (MCMC) steps.
The main advantages of the methodology are: i) it is sequential and
 on-line in nature; ii) it does
not require restrictive model assumptions such as Gaussian noise or
linear dynamics and observations; iii) it is 
parallelisable, so one could gain significant speed-up using appropriate
hardware (e.g.~GPUs, computing clusters) \citep{lee2010utility}; iv)
it is a well-studied principled method with an extensive literature
justifying its validity and theoretical properties, see e.g.~\citep{del2013mean,del2004feynman}.
So far SMC has been extremely successful in typically low to moderate dimensions \citep{doucet2001smc_practice}, but its
application in high dimensional settings has been very challenging
mainly due to the difficulty to perform IS efficiently in high dimensions
\citep{snyder2008obstacles}. Despite this challenge a few successful
high dimensional SMC implementations have appeared recently for applications
with discrete time signal dynamics  \citep{papadakis2010data,reich2013guided,van2010nonlinear,weare2009particle,bocquet2010beyond,chorin2010implicit,beskos2017stable}.

We will formulate the filtering problem with discrete time observations
and continuous time dynamics. This setup has appeared previously in
\citep{sarkka2008application,sarkka2016p} for signals corresponding to low dimensional	
stochastic differential equations (SDEs). The aim of this paper is
to provide a novel, accurate and more efficient SMC design when the
hidden signal is modelled by a SPDE with linear Gaussian observation.
To achieve this challenging task, the particle filter will use computational
tools that have been previously successful in similar high dimensional
problems, such as tempering \citep{kantas2014sequential} and pre-conditioned
Crank Nicholson MCMC steps \citep{hoang2014determining, cotter2013mcmc}.
Using such tools, we propose a particle algorithm that can be used
to approximate $\pi_{n}$ when the signal obeys the stochastic NSE
and the observations are linear with additive noise. On a general
level, the proposed algorithm has a similar structure to \citep{johansen2015block},
but here we additionally adopt the use of IS. We will provide a detailed
design of the necessary likehood informed importance proposals and
the MCMC moves used. We extend known IS techniques for SDEs (\citep{golightly2008bayesian,whitaker2016improved})
and MCMC moves for high dimensional problems (\citep{tierney1998note,hoang2014determining,cotter2013mcmc})
to make them applicable for filtering problems involving the stochastic NSE or
other dissipative SPDEs. In the context of particle filtering, our developments
leads to an SMC algorithm that performs effectively for the high
dimensional problem at hand using a moderate amount of particles.

{ The material presented in this paper can be viewed as an extension of some 
ideas in the authors' earlier work in \citep{kantas2014sequential}. In \citep{kantas2014sequential} 
we considered the deterministic NSE with more general observation equations. In the present paper the model for the signal contains additive noise 
and we assume linear observation schemes. This allows for the possibility of using likehood informed importance proposals
and the MCMC steps need to be designed to be invariant to a more complicated conditional law due to the SPDE dynamics.} The organisation of this paper is as follows: in Section \ref{sec:Background}
we present some background on the stochastic NSE and in
Section \ref{sec:The-stochastic-filtering} we formulate the filtering
problem of interest. In Section \ref{sec:Particle-filtering} we present the SMC
algorithm and in Section \ref{sec:Numerical-examples} we prsesent
a numerical case study that illustrates the performance and efficiency
of our method. Finally, in Section \ref{sec:Conclusions} we provide
some concluding remarks.

\section{Background on the Stochastic Navier-Stokes Equations\label{sec:Background}}

We present some background on the 2D stochastic NSE
defined on an appropriate separable Hilbert space. We restrict
the presentation to the case of periodic boundary conditions following
the treatment in \citep{ferrario1999stochastic}. This choice is motivated
mainly for convenience in exposition and for performing numerical approximations using Fast Fourier
Transforms (FFT).
The formulation and properties of the stochastic NSE can allow for
the more technically demanding Dirichlet conditions on a smooth boundary
\citep{flandoli1994dissipativity,hoang2014determining}. We stress that 
the subsequent particle filtering methodology is generic and does
not rely on the choice of boundary conditions. 

\subsection{Preliminaries}

Let the region of interest be the torus $\Omega:=[0,2\pi]^{2}$ with
$x=(x_{1},x_{2})\in\Omega$ being a point on the space. The quantity
of interest is a time-space varying velocity field $v:\Omega\times[0,T]\rightarrow\mathbb{R}^{2}$,
$v(x,t)=\left(v_{1}(x,t),v_{2}(x,t)\right)'$ {and $v(\cdot,0,t)=v(\cdot,2\pi,t)$ due to the periodic boundary conditions; here $\cdot^{'}$
denotes vector/matrix transpose. It is convenient
to work with the Fourier characterisation of the function space of interest:
\begin{equation}
H=\Big\{\,u=\sum_{k\in\mathbb{Z}^{2}\setminus\{0\}}\,u_{k}\,\psi_{k}(x)\,\,\big|\,\,u_{-k}=-\overline{u_{k}},\sum_{k\in\mathbb{Z}^{2}\setminus\{0\}}{|u_{k}|^{2}}<\infty\,\Big\},\label{eq:basis}
\end{equation}
using the following orthonormal basis functions for $H$, %s
\[
\psi_{k}(x)=\frac{1}{2\pi}\frac{k^{\perp}}{|k|}\,e^{i\,k\cdot x},\quad k\in\mathbb{Z}^{2}\backslash\{0\},\quad k^{\perp}:=(-k_{2},k_{1})'.
\]
The deterministic NSE is given by the following
the functional evolution equation 
\begin{equation}
dv+\nu Av\,dt+B(v,v)\,dt=f(t)\,dt,\quad v(0)\in H,\label{eq:sde}
\end{equation}
Following standard notation we denote $P:(L_{per}^{2}(\Omega))^{2}\rightarrow H$ for the Leray projector ($L_{\mathrm{per}}^{2}(\Omega)$ is the space of
squared-integrable periodic functions), $A:={P(-\Delta)=-\Delta}$ 
for the Stokes operator, $B(u,v)=P\big((u\cdot\nabla)v\big)$ for the convection mapping and $f\in L^{2}(0,T;H)$ for the forcing.}

One can introduce additive noise in the dynamics
in a standard manner. First, we define the upper half-plane of wavenumbers
\begin{align*}
\mathbb{Z}_{\uparrow}^{2}=\big\{ k=(k_{1},k_{2})\in\mathbb{Z}^{2}\backslash\{0\}:\:&k_{1}+k_{2}>0\big\} \\ &\cup\left\{ k=(k_{1},k_{2})\in\mathbb{Z}^{2}\backslash\{0\}:\:k_{1}+k_{2}=0,\,k_{1}>0\right\} .
\end{align*}
Let 
\[
Z_{k}(t)=Z_{k}^{re}(t)+i\,Z_{k}^{im}(t),\quad k\in\mathbb{Z}_{\uparrow}^{2}\ ,
\]
where $\{Z_{k}^{re},Z_{k}^{im}\}$ are (independent) standard Brownian
motions on $[0,T]$. In the spirit of \citep[Section 4.1]{da2008stochastic},
consider a covariance operator $Q$ such that $Q\psi_{k}=\sigma_{k}^{2}\psi_{k}$,
for $\sigma_{k}^{2}>0$, $\sigma_{-k}=\sigma_{k}$. Then, we can define
the $Q$-Wiener process as 
\begin{equation}
Q^{\frac{1}{2}}W(t):=\sum_{k\in\mathbb{Z}^{2}\setminus\{0\}}\sigma_{k}Z_{k}(t)\,\psi_{k}(x),\label{eq:QWiener}
\end{equation}
under the requirement {$Z_{-k}\equiv-\overline{Z_{k}}$, $k\in\mathbb{Z}_{\uparrow}^{2}$.}
Thus, we are working with a diagonal covariance matrix (w.r.t.~the
relevant basis of interest), though other choices could easily be
considered. We will also work under the scenario that $\sigma_{k}^{2}=O(|k|^{-2(1+\epsilon)})$,
for some $\epsilon>0$, so that $\sum_{k\in\mathbb{Z}^{2}\setminus\{0\}}\sigma_{k}^{2}<\infty$,
i.e.~$Q$ is trace-class operator. Finally, we will use $\mathbb{W}\left(\cdot\right)$
to denote the $Q$-Wiener measure on $[0,T]$.

Having introduced the random component, we are now interested in weak
solutions {$V=\big(V(t)\big)_{t\in[0,T]}$} of the functional SDE, 
\begin{equation}
dV(t)+\nu AV(t)\,dt+B(V(t),V(t))dt=f(t)\,dt+Q^{\frac{1}{2}}dW(t),\qquad V(0)=v_{0},\label{eq:spde}
\end{equation}
with the solution understood pathwise on the
probability space $(\boldsymbol{\Omega},\mathscr{F},(\mathscr{F}_{t})_{t\geq0},\mathbb{P})$.
More formally, following \citep{ferrario1999stochastic}, we define
the spaces
\[
\mathcal{V}_{s}:=\Big\{\,u=\sum_{k\in\mathbb{Z}^{2}\setminus\{0\}}\,u_{k}\,\psi_{k}(x)\,\,\big|\,\,u_{-k}=-\overline{u_{k}},\sum_{k\in\mathbb{Z}^{2}\backslash\{0\}}{|k|^{2s}|u_{k}|^{2}}<\infty\,\Big\},\quad s\in\mathbb{R}.
\]
Since the operator $Q^{1/2}$ is linear, bounded in $H$ and $\mathrm{Im}(Q^{1/2})\equiv\mathcal{V}_{1+\epsilon}$,
\citep[Theorem 6.1]{ferrario1999stochastic} implies that for $v_{0}\in\mathcal{V}_{1}$
and $f\in C([0,T];\mathcal{V}_{1})$, there exists a unique solution
for (\ref{eq:spde}) such that $V\in C\left([0,T];\mathcal{V}_{1}\right)$.
In \citep{ferrario1999stochastic,flandoli1994dissipativity} one may
also find more details on the existence of an invariant distribution, together with 
irreducibility and Feller properties of the corresponding Markov transition
kernel.

\subsection{Galerkin Projections and Computational Considerations}

Using the Fourier basis (\ref{eq:basis}), we can write the solution
as 
\begin{gather*}
V(t)=\sum_{k\in\mathbb{Z}^{2}\setminus\{0\}}u_{k}(t)\psi_{k}(x),\quad u_{-k}(t)\equiv-\overline{u_{k}(t)},\\ u_{k}(t)=\langle V(t),\psi_{k}\rangle=\int_{\Omega}V(t)\cdot\overline{\psi_{k}(x)}\,dx.
\end{gather*}
%
%with $u_{k}(t)=\langle v(\cdot,t),\psi_{k}\rangle=\int_{\Omega}v(x,t)\cdot\bar{\psi}_{k}(x)dx$
%($\bar{\cdot}$ denotes the complex conjugate). 
Hence, it is equivalent to consider the parameterisation of $V$ via $\left\{ u_{k}(t)\right\} _{k\in\mathbb{Z}^{2}\setminus\{0\}}$.
By taking the inner product with $\psi_{k}$
on both sides of (\ref{eq:spde}), it is straightforward to obtain  that the $u_{k}$'s obey the following
infinite-dimensional SDE 
\begin{align}
du_{k}(t)=-\nu|k|^{2}&u_{k}(t)\,dt \nonumber\\ &-\sum_{m,p\in\mathbb{Z}^{2}\setminus\{0\}}b_{k,m,p}u_{m}(t)u_{p}(t)dt+f_{k}(t)\,dt+\sigma_{k}dZ_{k}(t),\quad k\in\mathbb{Z}_{\uparrow}^{2},\label{eq:galerkinNS}
\end{align}
with 
\[
b_{k,m,p}=\left\langle B(\psi_{m},\psi_{p}),\psi_{k}\right\rangle ,\quad f_{k}(t)=\langle f(t),\psi_{k}\rangle.
\]
Recall that due to $V(t)$ being a real field $u_{-k}(t)\equiv-\overline{u_{k}(t)}$,
$k\in\mathbb{Z}_{\uparrow}^{2}$. %  , we will constraint the complex
%conjugate coefficients and split the domain by defining: 
%\[
%\mathbb{Z}_{\uparrow}^{2}=\left\{ k=(k_{1},k_{2})\in\mathbb{Z}^{2}\setminus\{0\}:\:k_{1}+k_{2}>0\right\} \cup\left\{ k=(k_{1},k_{2})\in\mathbb{Z}^{2}\setminus\{0\}:\:k_{1}+k_{2}=0,\,k_{1}>0\right\} \ .
%\]
%and then imposing that $u_{k}=-\overline{u_{-k}}$ for $k\in\{\mathbb{Z}^{2}\setminus\{0\}\}\setminus\mathbb{Z}_{\uparrow}^{2}$.
This parameterisation of $V$ is more convenient as it allows performing
inference on a vector (even if infinitely long), with coordinates evolving
according to an SDE. For numerical purposes one is forced to use Galerkin
discretisations, using projections of $V$ onto a finite Hilbert space
instead. Consider the set of wavenumbers in 
\[
\mathbb{L}=\left\{ k\in\mathbb{Z}_{\uparrow}^{2}:\left(k_{1}\lor k_{2}\right)\leq L\right\}, 
\]
for some integer $L>0$, and define the finite dimensional-subspace
$H_{L}$ via the projection $P_{L}:H\rightarrow H_{L}$ so that 
\[
P_{L}v=\sum_{k\in\mathbb{L}}\left\langle v,\psi_{k}\right\rangle \psi_{k}.
\]
Then, infering the Galerkin projection for $V$ corresponds
to inferring the vector $\left\{ u_{k}(t)\right\} _{k\in\mathbb{L}}$ that obeys the following finite-dimensional SDE 
\begin{equation}
du_{k}(t)=-\nu|k|^{2}u_{k}(t)-\sum_{m,p\in\mathbb{L}}b_{k,m,p}u_{m}(t)u_{p}(t)dt+f_{k}(t)\,dt+\sigma_{k}dZ_{k}(t),\quad k\in\mathbb{L}.\label{eq:finite_galerkin_SDE}
\end{equation}
This high dimensional SDE will provide an approximation for the infinite dimensional SPDE.
Such an inference problem is more standard, but is still challenging
due to the high dimensionality of $\mathbb{L}$ and the non-linearities
involved in the summation term of the drift function in (\ref{eq:finite_galerkin_SDE}).
Since (\ref{eq:finite_galerkin_SDE}) is only an approximation of
(\ref{eq:galerkinNS}),  it will induce a bias in the inferential procedure. In our paper, we do not study the size of this bias. Instead we concentrate our efforts on designing an algorithm
to approximate $\pi_{n}$ (in (\ref{eq:filter_intro})) that
is robust to \emph{mesh refinement}. This means our method should
perform well numerically when one increases $L$ (and, indeed, reducing the
bias in the numerical approximation of (\ref{eq:galerkinNS})). Naturally
this would be at the expense of adding computational effort at a moderate
amount, but this will depend on the particular numerical scheme used
to approximate the solution of (\ref{eq:finite_galerkin_SDE}). For
instance, for the FFT based numerical schemes used in Section \ref{sec:Numerical-examples}
the computational cost is $\mathcal{O}(L^{2}\log L)$.

\subsection{The Distribution of $v_{0}$}

We assume that the initial condition of $V$ is random and distributed
according to the following Gaussian process prior: 
\begin{equation}
\pi_{0}=\mathcal{N}(\mu,\beta^{2}A{}^{-\alpha}),\quad\alpha>2,\,\,\beta>0,\,\,\mu\in\mathcal{V}_{1},\label{eq:prior}
\end{equation}
with hyper-parameters $\alpha,\beta$ affecting the roughness and
magnitude of the initial vector field. This is a convenient but still
flexible enough choice of a prior; see \citep[Sections 2.3 and 4.1]{da2008stochastic}
for more details on Gaussian distributions on Hilbert spaces. Notice
that $\pi_{0}$ admits the Karhunen-Loève expansion 
\[
\pi_{0}=\mathcal{L}aw\left(\sum_{k\in\mathbb{Z}^{2}\setminus\{0\}}\left(\mu_{k}+\tfrac{\beta}{\sqrt{2}}\,|k|^{-\alpha}\,\text{\ensuremath{\xi}}_{k}\,\psi_{k}\right)\right),
\]
with $\mu_{k}=\left\langle \mu,\psi_{k}\right\rangle $, $k\in\mathbb{Z}_{\uparrow}^{2}$,
(so, necessarily $(\mu_{-k}=-\overline{\mu_{k}}$, $k\in\mathbb{Z}_{\uparrow}^{2}$)
and 
\[
\mbox{Re}(\xi_{k})\,,\,\mbox{Im}(\xi_{k})\,\stackrel{iid}{\sim}\mathcal{N}(0,1),\quad k\in\mathbb{Z}_{\uparrow}^{2}\ ;\quad\xi_{-k}=-\overline{{\xi}_{-k}},\quad k\in\mathbb{Z}_{\uparrow}^{2}.
\]
Since the covariance operator is determined via the Stokes operator
$A$, one can easily check that the choice $\alpha>2$ implies that
for $v_{0}\in\mathcal{V}_{1}$, $\pi_{0}$-a.s., thus the conditions
for existence of weak solution of (\ref{eq:spde}) in \citep[Theorem 6.1]{ferrario1999stochastic}
are satisfied a.s.~in the initial condition. Notice that sampling from $\pi_0$ is straightforward.

%Thus \emph{a-priori}, $v_{0}\sim\pi_{0}$ is equivalent to setting
%the Fourier coefficients as $u_{k}=\mu_{k}+\tfrac{\beta}{\sqrt{2}}\lambda_{k}{}^{-\frac{\alpha}{2}}\text{\ensuremath{\xi}}_{k}$
%for $k\in\mathbb{Z}_{\uparrow}^{2}$, which means they are assumed
%independent normally distributed with a particular rate of decay for
%their variances as $\left|k\right|$ increases. %
%\begin{comment}
%The requirement $\xi_{k}=-\bar{\xi}_{-k}$ is so that the inverse
%Fourier transform leads to a real function.
%\end{comment}
%\begin{comment}
%Then we truncate the initial random variable $v_{0}$, $A,B$ and
%the Browian motion $W_{t}$ by setting:
%\[
%v_{0}^{L}=P_{L}v_{0},\:A_{L}=P_{L}AP_{L},\:B_{L}\left(\cdot,\cdot\right)=P_{L}B\left(P_{L}\cdot,P_{L}\cdot\right),\:W_{t}^{L}=P_{L}W_{t}=\sum_{k\in\mathbb{L}}\sigma_{k}\,\psi_{k}(x)\,Z_{k,t}\ .
%\]
%can write Galerkin system in more detail here but maybe not worth
%it yet.
%\end{comment}

\section{The Stochastic Filtering Problem\label{sec:The-stochastic-filtering}}

In Section \ref{sec:Background} we defined the SPDE providing the
unknown signal, i.e.~the object we are interested in performing Bayesian
inference upon. In this section we present the non-linear filtering
problem in detail. We begin by discussing the observations. We
assume that the vector field $V$ is unknown, but generates a sequence
of noisy observations $\mathcal{Y}_{n}=\left(Y_{t_{1}},\ldots Y_{t_{n}}\right)$
at ordered discrete time instances $\left(t_{p}\right)_{p=1,\ldots,n}$
with $t_{n}<t_{n+1}<T$ for all $n$, with $Y_{t_{i}}\in\mathbb{R}^{d_{y}}$, 
for $d_y \ge 1$,
Each observation vector $Y_{t_{i}}$ is further assumed to originate
from the following observation equation 
\begin{equation}
Y_{t_{n}}=FV(t_{n})+\Xi_{n},\quad\Xi_{n}\sim\mathcal{N}(0,\Sigma),\label{eq:obs_eqn}
\end{equation}
where $F$ is a bounded linear operator $F:H\rightarrow\mathbb{R}^{d_{y}}$
and $\Sigma\in\mathbb{{R}}^{d_{y}\times d_{y}}$ is symmetric positive-definite.
One can then write the observation likelihood at instance $t_{n}$
as 
\[
%\frac{d\mathbb{P}}{dy}\big[\,Y_{t_{n}}\,|\,V(t_{n})\,\big]=
p(Y_{t_{n}}|V(t_{n}))=\frac{\exp\left(-\frac{1}{2}\left|\Sigma^{-\frac{1}{2}}\left(Y_{t_{n}}-FV(t_{n})\right)\right|^{2}\right)}{(2\pi)^{d_{y}/2}\left|\Sigma\right|^{1/2}}.
\]
Using a linear observation model is restrictive but it does include
typical observation schemes used in practice. We focus our attention
at the case when $Y_{t_{n}}$ is a noisy measurement of the velocity
field at different fixed stationary points $x_{l}\in\Omega$, $l=1,\ldots,p$.
This setting is often referred to as \emph{Eulerian data assimilation}. 
In particular we have that 
\begin{equation*}
F = (F_1',\ldots, F_{p}')',
\end{equation*}
with $F_l$ denoting 
a spatial average over a (typically small) region around  $x_{l}$, $l=1,\ldots,p$,
say $B_{x_{l}}(r)=\{x\in\Omega:|x-x_{l}|\le r\}$ for some radius $r>0$; that is, $F_l$
is the following integral operator 
\begin{equation}
F_lV(t)=\frac{1}{\left|B_{x_{l}}(r)\right|}\int_{B_{x_{l}}(r)}V(t,x)dx,\label{eq:observer}
\end{equation}
with $\left|B_{x_{l}}(r)\right|$ denoting the area of $B_{x_{l}}(r)$.
In what follows, other integral operators could also be similarly
used, such as $F_lV(t)=(\int_{\Omega}V(t,x)w_{x_{l}}(x)dx)/(\int_{\Omega}w_{x_{l}}(x)dx)$,
with $w_{x_{l}}\in L^{2}(\Omega)$ being appropriate weighting functions
that decay as $\left|x-x_{l}\right|$ grows.

Earlier in the introduction, the filtering problem was defined as
the task of computing the conditional distribution $\pi_{n}(\cdot)=\mathbb{P}\left[\left.V(t_{n})\in\cdot\right|\mathcal{Y}_{n}\right]$.
Due to the nature of the observations, it is clear  we are dealing
with a discrete time filtering problem. A particular challenge here (in common 
with other typical non-linear SPDEs)
is that the distribution of the associated Markov transition kernel, $\mathbb{P}\left[\left.V(t_{n})\in\cdot\right|V(t_{n-1})=v\right]$,
is intractable. Still, it is possible to simulate from the unconditional
dynamics of $V(t)$ given $V(t_{n-1})=v$ using standard time discretization
techniques. (The simulated path introduces a time discretisation
bias, but its effect is ignored in this paper.)

%At every $n$, assume one simulates the path of the SPDE between $t_{n-1}$
%and $t_{n}$ and consider that the object of inference is the complete
%SPDE path up to $t_{n}$. 
We aim to infer the following posterior
distribution, based on the \emph{continuous time signal} 
\begin{equation*}
\Pi_{n}(\cdot)=\mathbb{P}\left[\left.V^{n}\in\cdot\right|\mathcal{Y}_{n}\right],\quad V^{n}:=\left(V(t)\right)_{t\in[0,t_{n}]};
\end{equation*}
we also denote 
\begin{equation*}
V_{n-1}^{n}=(V(t))_{t\in(t_{n-1},t_{n}]}.
\end{equation*}
This data augmentation approach - when one applying importance sampling on continuous time -  has appeared in \citep{hoang2014determining}
for a related  problem and in \citep{sarkka2008application} for
filtering problems involving certain multivariate SDEs. We proceed
by writing the filtering recursion for $\Pi_{n}$.
We denote the law of $V$ in (\ref{eq:spde}) for the time interval
between $t_{n-1}$ and $t_{n}$ as 
\begin{equation*}
\mathbb{V}_{n-1}^{n}(\,\cdot\,|v):=\mathbb{P}\big[\left(V(t)\right)_{t\in(t_{n-1},t_{n}]}\in\,\cdot\,\big|V(t_{n-1})=v\big].
\end{equation*}
Then, one may use Bayes rule to write $\Pi_{n}$ recursively as
\begin{equation}
\frac{d\Pi_{n}}{d\left(\Pi_{n-1}\otimes\mathbb{V}_{n-1}^{n}\right)}\left(V^{n}\right)=\frac{p\left(Y_{t_{n}}|V(t_{n})\right)}{p(Y_{t_{n}}|\mathcal{Y}_{n-1})},\label{eq:bayes_filter}
\end{equation}
where $p(Y_{t_{n}}|\mathcal{Y}_{n-1})=\int p\left(Y_{t_{n}}|V(t_{n})\right)\left[\Pi_{n-1}\otimes\mathbb{V}_{n-1}^{n}\right]\left(dV^{n}\right)$.

In addition, one can attempt to propose paths from an appropriate SPDE different
from (\ref{eq:spde}), say
\begin{align}
d\tilde{V}(t)+\nu A\tilde{{V}}(t)dt+&B(\tilde{{V}}(t),\tilde{{V}}(t))dt \nonumber \\ &=Q^{\frac{1}{2}}g(t,\tilde{V}(t))dt+f(t)\,dt+Q^{\frac{1}{2}}{dW(t)},\qquad t\in(t_{n-1},t_{n}],\label{eq:proposal}
\end{align}
where $g:[0,T]\times H\mapsto H$ and $Q^{\frac{1}{2}}W_{t}$ is a $Q$-Wiener process on $(t_{n-1},t_{n}]$.
We define
\begin{equation*}
\mathbb{Q}_{n-1}^{n}(\,\cdot\,|v):=\mathbb{P}\big[{\tilde{V}_{n-1}^{n}}\in\,\cdot\,\big|\tilde{V}(t_{n-1})=v\big].
\end{equation*}
One needs to ensure that the change of drift $g$ is appropriately chosen
so that a Girsanov theorem holds and $\mathbb{V}_{n-1}^{n}(\cdot|v)$
is absolutely continuous with respect to $\mathbb{Q}_{n-1}^{n}(\cdot|v)$ for all relevant 
$v$,
with the recursion in (\ref{eq:bayes_filter}) becoming
\begin{equation}
\label{eq:ISweight}
\frac{d\Pi_{n}}{d\left(\Pi_{n-1}\otimes\mathbb{Q}_{n-1}^{n}\right)}(V^{n-1},\tilde{V}_{n-1}^{n})\propto p(Y_{t_{n}}|\tilde{V}(t_{n}))\cdot\frac{d\mathbb{V}_{n-1}^{n}}{d\mathbb{Q}_{n-1}^{n}}(\tilde{V}_{n-1}^{n}|{V(t_{n-1}})).
\end{equation}
Here $(V^{n-1},\tilde{V}_{n-1}^{n})$ are assumed to be typical elements of the sample space of either of 
the two probability measures above 
(e.g. all such paths are assumed to possess relevant continuity properties at $t_{n-1}$).

In the context of particle filtering and IS one aims to design $g$
in a way that the proposed trajectories are in locations where $\Pi_{n}$
is higher. This in turn implies that the importance weights in (\ref{eq:ISweight})
will exhibit much less variance than the ones from the prior signal dynamics,
hence the design of $g$ is critical for generating effective Monte Carlo approximations.

\section{Particle Filtering\label{sec:Particle-filtering}}

We are interested in approximating the distribution $\Pi_{n}$ using
a particle filter approach. We present in Algorithm~\ref{alg:smc-1}
a naive particle filter algorithm that provides the particle approximations:
\begin{equation*}
\Pi_{n}^{N}=\sum_{j=1}^{N}\mathscr{W}_{n}^{i}\delta_{V^{i}}\,\mbox{ or }\,\bar{\Pi}_{n}^{N}=\frac{1}{N}\sum_{j=1}^{N}\delta_{\bar{V}^{i}}.
\end{equation*}
Such a particle filter will be typically overwhelmed by the dimensionality
of the problem and will not be able to  provide accurate solutions with a moderate
computational cost. When $g=0$ in (\ref{eq:proposal}), the algorithm
corresponds to a standard bootstrap particle filter. For the latter,
it is well known in the literature (\citep{bocquet2010beyond,snyder2008obstacles})
that it exhibits weight degeneracy in the presence of
large dissimilarity between $\Pi_{n-1}\otimes\mathbb{V}_{n-1}^{n}$
and $\Pi_{n}$, which can be caused in our context by the high dimensionality
of the state space and the complexity of the SPDE dynamics. When $g$
is well designed then the particles can be guided in areas of larger
importance weights and the algorithmic performance can be considerably improved,
but this modification may still not be sufficient for obtaining a robust and efficient algorithm.

\begin{algorithm}[h]
\begin{itemize}
\item Initialise $V_{0}^{i}\sim\pi_{0}$, $1\le i\le N$. 
\item For $n\geq1$
\begin{enumerate}
\item For $i=1,\ldots N$: sample independently 
\[
\tilde{V}_{n-1}^{n,i}\sim\mathbb{Q}_{n-1}^{n}(\,\cdot\,|{V(t_{n-1})^{i}}).
\]
\item For $i=1,\ldots N$: compute importance weights 
\[
\mathscr{W}_{n}^{i}\propto p(Y_{t_{n}}|\tilde{V}(t_{n})^{i})\cdot\frac{d\mathbb{V}_{n-1}^{n}}{d\mathbb{Q}_{n-1}^{n}}({V}_{n-1}^{n,i}|{V(t_{n-1})^{i}}),\quad\text{s.t. }\sum_{i=1}^{N}\mathscr{W}_{n}^{i}=1.
\]
\item For $i=1,\ldots N$: resample 
\[
V^{n,i}\sim\sum_{j=1}^{N}\mathbf{\mathscr{W}}_{n}^{j}\,\delta_{(V^{n-1,j},\tilde{V}_{n-1}^{n,j})}(\,\cdot\,).
\]
\end{enumerate}
\end{itemize}
\caption{A naive particle filter}

\label{alg:smc-1} 
\end{algorithm}

In the remainder of this section, we will discuss how to improve upon this
first attempt to tackle the high-dimensional filtering problem at hand using the following
ingredients: (i) specifying a particular form of $g$ in (\ref{eq:proposal})
that results in gains of efficiency, (ii) using adaptive tempering,
and (iii) MCMC moves. Guided proposals and tempering are employed
to bridge the dissimilarity between $\Pi_{n-1}\otimes\mathbb{V}_{n-1}^{n}$
and $\Pi_{n}$. The MCMC steps are required for injecting additional
diversity in the particle population, which would otherwise diminish
gradually due to successive resampling and tempering steps. The method
is summarised in Algorithm \ref{alg:smc}. In the following subsections
we  explain in detail our implementation of (i)-(iii) mentioned above.

\begin{comment}
XXX we could add a generic bound result here and mention that despite
this algorithm being provably convergent we want to make it more efficient
XXX

IS using likelihood informed or guided proposals that sample from
a SPDE like (\ref{eq:proposal}) 
\end{comment}

\subsection{Likelihood-Informed Proposals}

In the importance weight of (\ref{eq:ISweight}) we are
using a Girsanov Theorem and assume absolute continuity between SPDEs (\ref{eq:proposal})
and (\ref{eq:spde}) when started at the same position. Under the assumption
\begin{equation}
\mathbb{P}\Big[\int_{0}^{T}\left\Vert g(t,V(t))\right\Vert ^{2}dt<\infty\Big]=1,\label{eq:liptser-shiraev}
\end{equation}
%
\begin{comment}
Goldys-Maslovski 06, Ferrario 08 (Liptser-Shiryaev conditions) 
\end{comment}
absolute continuity indeed holds and we have Radon-Nikodym derivative
\begin{align*}
\log&\frac{d\mathbb{V}_{n-1}^{n}}{d\mathbb{Q}_{n-1}^{n}}({\tilde{V}_{n-1}^{n}}|V^{n-1}(t_{n-1})) \\ &
 \qquad\qquad\qquad=
-\int_{t_{n-1}}^{t_{n}}\langle Q^{\frac{1}{2}} g(t,{\tilde{V}(t)}),Q^{\frac{1}{2}}dW(t)\rangle_{0}-\tfrac{1}{2}\int_{t_{n-1}}^{t_{n}}\left\Vert  Q^{\frac{1}{2}}g(t,{\tilde{V}(t)})\right\Vert _{0}^{2}dt,
\end{align*}
where 
\[
\langle u,v\rangle_{0}:=\langle Q^{-\frac{1}{2}}u,Q^{-\frac{1}{2}}v\rangle\equiv\sum_{k\in\mathbb{Z}^{2}\setminus\{0\}}\tfrac{1}{\sigma_{k}^{2}}\,\langle u,\psi_{k}\rangle\langle v,\psi_{k}\rangle;
\]
see \citep[Theorem 10.14]{da2008stochastic} and \citep[Lemma 10.15]{da2008stochastic}
for details. It remains to provide an effective design for $g$. One can
use proposals developed for problems whereby a finite-dimensional SDE generates linear
Gaussian observations and one is interested to perform a similar IS
method, see e.g.~\citep{golightly2008bayesian,whitaker2016improved,papaspiliopoulos2012importance,papaspiliopoulos2013data,van2016bayesian}.
In this paper we use the proposal employed in \citep{golightly2008bayesian}
and set 
\begin{equation}
g(t,V(t))=Q^{\frac{1}{2}}F^{*}(\Sigma+(t_{n}-t)FQF^{*})^{-1}(Y_{t_{n}}-FV(t)),\quad t\in(t_{n-1},t_{n}],\label{eq:guiding function}
\end{equation}
where $F^{*}$ denotes the adjoint of $F$. The guiding function $g$
could be interpreted as an one-step Euler approximation of the $h$-tranform
needed to evolve $V(t)$ conditional on the acquired observation $Y_{t_{n}}$
within the interval $(t_{n-1},t_{n}]$. It is not hard to verify (\ref{eq:liptser-shiraev})
for this choice of $g$. Since $\Sigma$,$Q$ are invertible then
$(\Sigma+(t_{n}-t)FQF^{*})^{-1}$ exists via the Sherman-Morrison-Woodbury
identity and $Q^{\frac{1}{2}}F^{*}(\Sigma+(t_{n}-t)FQF^{*})^{-1}$
is a bounded linear operator. Then (\ref{eq:liptser-shiraev}) holds
from \citep[Proposition 10.18]{da2008stochastic} and \citep[Proposition 2.4.9]{kuksin2012mathematics}
that imply that there exists a $\delta>0$ such that 
\begin{equation*}
\sup_{t\in[0,T]}\mathbb{E}\left[\exp\left(\delta\left\Vert g(t,V(t))\right\Vert ^{2}\right)\right]<\infty,
\end{equation*}
which implies (\ref{eq:liptser-shiraev}).

For the finite-dimensional SDE case more elaborate guiding functions
can be found in \citep{whitaker2016improved,van2016bayesian} and
some of these could be potentially extended so that they can be used
in the SPDE setting instead of (\ref{eq:guiding function}). The advantage of using $g$
in (\ref{eq:guiding function}) is that it provides a simple functional
and can perform well for problems where $t_{n}-t_{n-1}$ is of moderate
length, as also confirmed in the numerical examples of Section \ref{sec:Numerical-examples}.

\subsection{Bridging $\Pi_{n-1}$ and $\Pi_{n}$ With Adaptive Tempering }		

{ Guided proposals aim to bridge the dissimilarity between $\Pi_{n-1}\otimes\mathbb{V}_{n-1}^{n}$
and $\Pi_{n}$ by considering a Bayesian update from $\Pi_{n-1}\otimes\mathbb{Q}_{n-1}^{n}$
to $\Pi_{n}$}. In a high-dimensional setting, even using well-designed likelihood-informed proposals is not
sufficient to bridge the dissimilarity between the informed proposal $\Pi_{n-1}\otimes\mathbb{Q}_{n-1}^{n}$
and the target $\Pi_{n}$. As a result the importance weights could still degenerate. 
{To avoid this more effort is required. One possibility is to allow for a progressive 
update via a sequence of intermediate artificial distributions between $\Pi_{n-1}\otimes\mathbb{Q}_{n-1}^{n}$
to $\Pi_{n}$, which we will denote as $\Pi_{n,l}$ with $l=1,\ldots,\tau_n$ and require that $\Pi_{n,0}=\Pi_{n-1}\otimes\mathbb{Q}_{n-1}^{n}$
and $\Pi_{n,\tau_n}=\Pi_{n}$. This is a well known strategy to improve particle filters; see \citep{oudjane2000progressive,godsill2001improvement} for some 
early works in this direction for low dimensional problems.

To construct the sequence we will use a standard tempering scheme \citep{neal2001annealed}}. Each $\Pi_{n,l}$ can be defined using 
\begin{equation}
\frac{d\Pi_{n,l}}{d\left(\Pi_{n-1}\otimes\mathbb{Q}_{n-1}^{n}\right)}({V^{n-1},\tilde{V}_{n-1}^{n}})\propto\left(\frac{d\mathbb{V}_{n-1}^{n}}{d\mathbb{Q}_{n-1}^{n}}({\tilde{V}_{n-1}^{n}})\cdot
p\left(Y_{t_{n}}|{\tilde{V}}(t_{n})\right)\right){}^{\phi_{n,l}},\label{eq:Pi_n_l}
\end{equation}
for inverse temperatures $0=\phi_{n,0}<\phi_{n,1}<\cdots<\phi_{n,\tau_{n}}=1$.
%Note that by construction we have $\Pi_{n,\tau_n}\equiv \Pi_n$ as required and a 
{ Note that each $\Pi_{n,l}$ is defined on the same state space of $V^n$ and there are
no natural stochastic dynamics connecting each $\Pi_{n,l}$. 
As a result we will follow the framework of \citep{del2006sequential,chopin2002sequential} and 
use artificial dynamics provided by a MCMC transition kernel that is invariant to $\Pi_{n,l}$. 
The details are provided in the next section. Using these MCMC proposals will result in the weights at iteration $(n,l)$  being
$\mathscr{W}_{n,l}^{j}\propto\frac{d\Pi_{n,l}}{d\Pi_{n,l-1}}$, which depends on $\phi_{n,l}-\phi_{n,l-1}$ and the proposed $V^n$ from the MCMC kernel.
}

The main issue that needs addressing for this scheme to be successful is how to determine the temperatures $\phi_{n,l}$ 
and their number $\tau_{n}$. We propose to set these on-the-fly using
an adaptive procedure introduced in \citep{jasra2011inference}.
Assume we are at the $n$-th
step of the algorithm, have completed $l-1$ tempering steps, and
we have equally weighted particles. The next temperature is determined
by expressing the weights as a function of $\phi$ 
\begin{gather}
\mathscr{W}_{n,l}^{j}(\phi)\propto\left(\frac{d\mathbb{{V}}_{n-1}^{n}}{d\mathbb{Q}_{n-1}^{n}}({\tilde{V}_{n-1}^{n,j}})
p\left(Y_{t_{n}}|{\tilde{V}}(t_{n})^{j}\right)\right){}^{\phi-\phi_{n,l-1}},\quad\phi_{n,l-1}<\phi\leq1,\label{eq:weight}\\ \sum_{i=1}^{N}\mathscr{W}_{n,l}^{i}(\phi)=1, \nonumber
\end{gather}
and determining $\phi_{n,l}$ via a requirement based on a quality
criterion for the particle population. We use here the Effective
Sample Size (ESS), and set
\begin{equation}
\phi_{n,l}=\inf\Big\{\,\phi\in(\phi_{n,l-1},1]:ESS_{n,l}(\phi):=\tfrac{1}{\sum_{j=1}^{N}\{\mathscr{W}_{n,l}^{j}(\phi)\}^{2}}\le\alpha N\Big\},\label{eq:ESS_phi}
\end{equation}
(under the convention that $\inf \varnothing =1$) with a user-specified fraction $\alpha\in(0,1)$. Equation (\ref{eq:ESS_phi}) can be easily
solved numerically using for instance a standard bisection method. This approach 
leads to a particle approximation for $\Pi_{n,l}$, say  $$\Pi_{n,l}^{N}=\sum_{i=1}^{N}\mathscr{W}_{n,l}^{i}(\phi_{n,l})\delta_{V^{n,i}};$$
we then propose to resample from $\Pi_{n,l}^{N}$ so that one
ends up with equally weighted particles.

{The adaptive tempering procedure is presented in step 3 of Algorithm \ref{alg:smc}. 
In steps 3(a)-3(c), \eqref{eq:weight}-\eqref{eq:ESS_phi} are followed by resampling and MCMC steps and the steps are iterated
until $\phi_{n,l}=1$. The MCMC dynamics are denoted by $\mathcal{K}_{n,l}^{m}$ and will be discussed below.
For every $n$, the output of step 4 of Algorithm \ref{alg:smc}
provides a particle approximation
$\Pi_{n}^{N}=\frac{1}{N}\sum_{i=1}^{N}\delta_{V^{n,i}}$ targeting $\Pi_{n}$. The interesting feature of this algorithm is that when
moving from $\Pi_{n-1}$ to $\Pi_{n}$, it does not required a user-specified
intermediate sequence of target distributions $\left(\Pi_{n,l}\right)_{l=0,\ldots,\tau_{n}}$, but 
these are adaptively set according to the locations of the particles and \eqref{eq:ESS_phi}. 
The number of steps required, $\tau_n$, 
will be determined according to the difficulty in assimilating $Y_{t_n}$.}

\begin{rem}
The convergence of Algorithm \ref{alg:smc} has been studied in \citep{beskos2016convergence,giraud2017nonasymptotic}.
\end{rem}
{
\begin{rem}
%One may avoid resampling if at the last temperature $\phi_{n,\tau_{n}}=1$ and $ESS_{n,l+1}(\phi_{n,\tau_{n}})>\alpha N$.
%Then 
In Algorithm \ref{alg:smc} for simplicity we always resample once $\phi_{n,\tau_{n}}=1$. This can be avoided, but then in the next time-step of the algorithm one should use 
\[
\mathscr{W}_{n+1,0}^{j}(\phi)=\mathscr{W}_{n,\tau_{n}}^{j}\cdot\left(\frac{d\mathbb{V}_{n}^{n+1}}{d\mathbb{Q}_{n}^{n+1}}(\tilde{V}_{n}^{n+1,j})\cdot p(Y_{t_{n+1}}|\tilde{V}(t_{n+1})^{j})\right){}^{\phi}.
\]
\end{rem}
}

\subsection{Adding Particle Diversity With MCMC kernels}

Successive resampling due to the tempering steps leads to sample
impoverishment unless the method re-injects sampling diversity. To achieve
this, we propose using a small number of iterations from a MCMC procedure
that leaves $\Pi_{n,l}$ invariant. { This is not the only possible choice, but it does lead to a
a simple weight expression seen above; see \citep{del2006sequential} for extensions and more details.} 
We use a particular MCMC design similar to \citep{hoang2014determining}
that is well defined on function spaces (based on theory for MCMC on general state spaces
\citep{tierney1998note}). The design is often referred to as preconditioned Crank-Nicolson, 
abbreviated here to pCN; see \citep{stuart2010inverse,cotter2013mcmc}
for a detailed review. 

{ We begin with a basic description of the pCN scheme for a given target distribution $\Pi$; for simplicity we will drop the subscripts $n,l$ here.}  We will denote
the { one step} Markov probability kernel obtained from the MCMC procedure as 
\begin{equation}
\mathcal{K}\left[\left.V'\in\cdot\right|V\right]=\alpha\left(V,V'\right)\mathcal{Q}\left[\left.V'\in\cdot\right|V\right]+\delta_{V}(\cdot)\left(1-\int\alpha\left(V,V'\right)\mathcal{Q}\left[\left.dV'\right|V\right]\right)\label{eq:K_n_l},
\end{equation}
with $\mathcal{Q}$ denoting the proposal kernel and $\alpha$ 
the acceptance probability in a standard Metropolis-Hastings framework.
%The Markov kernel $\mathcal{K}_{n,l}$ uses another Markov kernel
%$\mathcal{Q}$ to generate a proposal $V'$, which is accepted as
%a sample from $\mathcal{K}_{n,l}$ with probability $\alpha_{n,l}$
%otherwise the process stays at its current state $V$.
Let $\Lambda$ be
a probability measure that is absolutely continuous with respect to
$\Pi$ with Radon-Nikodym derivative 
\[\frac{d\Pi}{d\Lambda}(V)=:\vartheta(V).\]
Similar to \citep{stuart2010inverse,cotter2013mcmc,hoang2014determining}
we specify the proposal kernel $\mathcal{Q}$ to satisfy detailed balance with respect to 
$\Lambda$, i.e.~$\mathcal{Q}(\left.dV'\right|V)\Lambda(dV)=\mathcal{Q}(\left.dV\right|V')\Lambda(dV')$.
Then, using 
\[\alpha(V,V')=1\land\frac{\vartheta(V')}{\vartheta(V)}\]
provides a kernel $\mathcal{K}$ which is $\Pi$-invariant (by
\citep[Theorem 2]{tierney1998note}).

{ Next we discuss implementing the pCN design for our problem.
At iteration $(n,l)$ the target distribution for the MCMC kernels is $\Pi_{n,l}$, so 
let $\mathcal{K}_{n,l}$, $\mathcal{Q}_{n,l}$ and $\alpha_{n,l}$ 
denote the corresponding MCMC kernel, proposal and acceptance ratio respectively.
Note that the state space of $\Pi_{n,l}$ is the space of paths $V^n$, which is growing with each observation time $n$.} 
We stress that for the purpose of particle filtering we are
mainly interested on the invariance property of $\mathcal{K}_{n,l}$ {(to $\Pi_{n,l}$)}
and not necessarily its ergodic properties on the full space. With this
in mind $\mathcal{Q}_{{n,l}}$ can be a Markov kernel that generates proposals $V'$
with $V_{s}'=V_{s}$ for $s\leq t_{n-1}$. This allows for on-line
computation at each $n,l$. At the same time reversibility holds
as Proposition 1 and Theorem 2 in \citep{tierney1998note} still hold
for such proposals. From a practical perspective, we are adding noise
to the path of the hidden signal only within $(t_{n-1},t_{n}]$.

{Then, we need to specify $\Lambda_{n}$
and $\mathcal{Q}_{n,l}$. Recall that for a fixed $n$ the state space of each $\Pi_{n,l}$ is the same for different $l$, so $\Lambda_{n}$
needs not vary with $l$.
One possibility is to let  $\Lambda_{n}=\Pi_{n-1}\otimes\mathbb{Q}_{n-1}^{n}$ and suppose $V_{n-1}^{n}=V_{n-1}^{n}(W)$}
with $W$ being the driving noise that generated $V_{n-1}^{n}$. Note that
we can assume than $W(t_{n-1})=0$ without loss of generality, since the $V$-path uses the increments of $W$. 
Suppose also that both $V_{n-1}^{n}$ 
and $W$ are stored in the computer's memory and 
so that 
\begin{equation*}
\vartheta_{n,l}({V^{n-1},\tilde{V}_{n-1}^{n}})=\frac{d\Pi_{n,l}}{d\Lambda_{n}} ({V^{n-1},\tilde{V}_{n-1}^{n}})=
\left(\frac{d\mathbb{V}_{n-1}^{n}}{d\mathbb{Q}_{n-1}^{n}}({\tilde{V}_{n-1}^n})\,p\left(Y_{t_{n}}|{\tilde{V}^n}(t_{n})\right)\right){}^{\phi_{n,l}}.
\end{equation*}
To simulate from a $\Lambda_n$-preserving proposal one first generates a new noise
sample $W'$ 
\begin{equation}
W(s)'=\rho\,W(s)+\sqrt{1-\rho^{2}}\,\xi(s),\quad t_{n-1}<s\leq t_{n},\quad \xi\sim\mathbb{W},\label{eq:W_mcmc_proposal}
\end{equation}
where $W(s)$ is the noise driving $V$ and {$\mathbb{W}$ is the Q-Wiener measure}. To return to the original space, 
we use the new noise  $W'$ to solve for $V'$ in (\ref{eq:proposal}). A standard calculation
can show that $W'\sim\mathbb{W}$, which in turn implies that for
the part of the proposal $V'$ in $(t_{n-1},t_{n}]$, $ (V_{n-1}^{n} )'\sim\mathbb{Q}_{n-1}^{n}$
holds. Reversibility with respect to $\Lambda$ is ensured using a simple conditioning
and marginalization argument. 

{In Algorithm \ref{alg:smc} we use $m$ iterations of \eqref{eq:K_n_l} with  $\mathcal{Q}_{n,l}$ specified as above.
The corresponding m-iterate of the MCMC transition kernel is denoted as $\mathcal{K}_{n,l}^{m}$ and
is presented in Algorithm \ref{alg:mcmc} in an algorithmic form. To simplify exposition, in  Algorithm \ref{alg:smc}
for each iteration $(n,l)$ the simulated tempered path $\tilde{V}_{n-1}^{n}$ for particle $i$ is denoted as $X_{n,l}^{i}$ and 
the MCMC mutation is presented jointly with resampling in step 3 (c) ii.}

\begin{algorithm}[h]
\begin{itemize}
\item At $n=0$. For $i=1,\ldots,N$,  sample i.i.d. $V_{0}^{i}\sim\pi_{0}$, and set $\mathscr{W}_{0}^{i}=1/N$. 
\item At time $n\ge1$. %For $i=1,\ldots,N$:
\begin{enumerate}
\item For $i=1,\ldots N$: sample independently 
\begin{equation*}
X_{n}^{i}\sim\mathbb{Q}_{n-1}^{n}(\,\cdot\,|V^{n-1,i}(t_{n-1}))
\end{equation*}
\item Set $l=0$, $X_{n,0}^{i}=X_{n}^{i}$, $\Pi_{n,0}=\Pi_{n-1}\otimes\mathbb{Q}_{n-1}^{n}$,
$\phi_{n,0}=0$. 
\item While $\phi_{n,l}<1$ 
\begin{enumerate}
\item Set $l\leftarrow l+1$ 
\item Specify $\Pi_{n,l}$, $\phi_{n,l}$ based on the $\mathrm{ESS}$ computation
in (\ref{eq:ESS_phi}) 
\item For $i=1,\ldots N$ 
\begin{enumerate}
\item Compute weights $\mathscr{W}_{n,l}^{i}$ as in (\ref{eq:weight}) 
\item Resample and move particles: 
\begin{equation*}
X_{n,l}^{i}\stackrel{i.i.d.}{\sim}\sum_{j=1}^{N}\frac{\mathscr{W}_{n,l}^{j}}{\sum_{k=1}^{N}\mathscr{W}_{n,l}^{k}}\,\mathcal{K}_{n,l}^{m}(\,\cdot\,|\,X_{n,l-1}^{j})
\end{equation*}
%
%\item Set $X_{n,l}^{i}=\bar{X}_{n,l}^{i}$ 
\end{enumerate}
\end{enumerate}
\item If $\phi_{n,l}=1$ return $V^{n,i}=(V^{n-1,i},X_{n,l}^{i})$, $\tau_{n}=l$;
otherwise go back to Step 3. 
\end{enumerate}
\end{itemize}
\caption{Adaptive Particle Filtering Algorithm %here we denote $(V(t)^{i})_{t\in(t_{n-1},t_{n}]}$
%as $X_{n}^{i}$ or $X_{n,l}^{i}$.
}

\label{alg:smc} 
\end{algorithm}

\subsubsection{Extensions\label{sec:MCMCextensions}}	
Firstly, similarly with \citep{doucet2006efficient}  one can extend the proposals by reducing the lower bound on the
time we start adding noise (here $t_{n-1}$). This could be made smaller
and this can be beneficial in terms of adding diversity, but for the
sake of simplicity we do not pursue this further.

{
It is important to note that $\mathcal{K}_{n,l}$ is based on adapting a very basic version of pCN-MCMC as 
outlined in \citep{stuart2010inverse,cotter2013mcmc,hoang2014determining}. There, typically $\Lambda$ is chosen to be 
a Gaussian measure that concides with a pre-chosen prior for a static Bayesian inference problem.
The resulting MCMC kernel often exhibits slow mixing properties.
This can be addressed by allowing a few selected coordinates be proposed from a kernel invariant to a Gaussian approximation of
the posterior distribution. The remaining coordinates are sampled as before 
(using kernels invariant to the prior), so that the scheme is valid for arbitrary dimensions.
This results in more advanced pCN-MCMC algorithms with likelihood informed proposals for $\mathcal{Q}_{n,l}$
such as the ones described in \citep{cui2016dimension,law2014proposals}. In the context of SMC 
one has the added benefit of using particle approximations
for the mean and covariance to construct likelihood informed proposals for $\mathcal{Q}_{n,l}$ and
this results to a simple and effective approach as illustrated in \citep{kantas2014sequential, beskos2017multilevel}.

A natural question to pose is how these ideas can be extended to construct more efficient $\mathcal{K}_{n,l}$.
Note that the filtering problem is more complicated as the variables of interest are SPDE paths.
Still more advanced proposals can be implemented after a change of measure. 
For the MCMC above we chose $\Lambda_n=\Pi_{n-1}\otimes\mathbb{Q}_{n-1}^n$. 
This choice was because of its simplicity in implementation and its effectiveness in the numerical examples we considered, where
the MCMC kernel in Algorithm \ref{alg:mcmc} mixed well. 
When facing harder problems, one can extend the construction of $\Lambda_n$ and use instead of $\mathbb{Q}^n_{n-1}$ any measure that admits a 
Radon-Nicodym derivative w.r.t it. For example one could use instead of \eqref{eq:W_mcmc_proposal} a proposal like
\begin{equation}
V(s)'=\rho\,V(s)+\sqrt{1-\rho^{2}}\,W'(s),\quad t_{n-1}<s\leq t_{n},\quad W'\sim{\mathbb{W}},\label{eq:V_mcmc_proposal}
\end{equation}
with $\vartheta_{n,l}=\frac{d\Pi_{n,l}}{d(\Pi_{n-1}\otimes\mathbb{W}_{n-1}^n)}$, where the 
Girsanov tranformation between $(t_{n-1},t_n]$ can be established rigourously as in \citep[Propositions 4.1 and 4.2]{chang1996large}. 
This construction is an alternative to Algorithm \ref{alg:mcmc} that is more ameanable to extensions along the 
lines of \citep{kantas2014sequential, beskos2017multilevel} as the reference measure is Gaussian. 
To follow \citep{kantas2014sequential, beskos2017multilevel} one should  use a Gaussian measure whose covariance operator should
take into account the likelihood for low frequencies. This means one should use in \eqref{eq:V_mcmc_proposal} 
a different Gaussian measure in \eqref{eq:V_mcmc_proposal} than $\mathbb{W}$,
which is identical to  $\mathbb{W}$ for high $|k|$ and for low $|k|$ the diffusion constants are computed from 
particle approximations for the posterior mean and covariance (given $\mathcal{Y}_n$) of a sequence $(W_{t_i}; t_i=t_{n-1},\ldots,t_{n})$
obtained just before the MCMC mutation.}

\begin{algorithm}[h]
\begin{itemize}
\item Initialise: set $V^{(0)}=X_{n,l}^{i}$ and let $W^{(0)}=W_{n,l}^{i}$
be the Wiener process generating $X_{n,l}^{i}$.
\item For $k=1,\ldots,m$: let $V=V^{(k-1)}$, $W=W^{(k-1)}$.
\begin{itemize}
%\item \textbf{if} $n=1$ only, obtain a sample for initial condition $V'_{t_{0}}=\mu+\rho_{0}\,\left(V_{t_{0}}-\mu\right)+\sqrt{1-\rho_{0}^{2}}\,\tilde{V}{}_{t_{0}}\quad\tilde{V}{}_{t_{0}}\sim\mathcal{N}(\mu,\beta^{2}A{}^{-\alpha})$
\item Sample a new noise
\begin{equation*}
W(s)'=\rho\,W(s)+\sqrt{1-\rho^{2}}\,\xi(s),\quad s\in(t_{n-1},t_{n}],\quad \xi\sim\mathbb{W}.
\end{equation*}
\item Obtain solution of SPDE \eqref{eq:proposal} with $W'$ the driving noise, i.e.
\begin{gather*}
dV'(s)=\left(-\nu AV'(s)-B(V'(s),V'(s))\right)dt+Q^{\frac{1}{2}}g(s,V'(s))ds+Q^{\frac{1}{2}}dW'(s), \\ t\in(t_{n-1},t_{n}].
\end{gather*}
\item Compute acceptance ratio
\begin{equation*}
\alpha_{n,l}=1\wedge\frac{\left(\frac{d\mathbb{V}^{n}}{d\mathbb{Q}^{n}}(V')p\left(Y_{t_{n}}|V'(t_{n})\right)\right){}^{\phi_{n,l}}}{\left(\frac{d\mathbb{V}^{n}}{d\mathbb{Q}^{n}}(V)p\left(Y_{t_{n}}|V(t_{n})\right)\right){}^{\phi_{n,l}}}.
\end{equation*}
\item With probability $\alpha_{n,l}$ set $V^{(k)}=V'$, $W^{(k)}=W'$; otherwise
reject proposal, set $V^{(k)}=V$, $W^{(k)}=W$.
\end{itemize}
\item Return $\bar{X}_{n,l}^{i}=V^{(k)}$ and $\bar{W}_{n,l}^{i}=W^{(k)}$.
\end{itemize}
\caption{An MCMC Procedure for $\bar{X}_{n,l}^{i}\sim\mathcal{K}_{n,l}^{m}(\left.\cdot\right|X_{n,l}^{i})$.}
\label{alg:mcmc} 
\end{algorithm}

\section{Numerical examples\label{sec:Numerical-examples}}

We solve SPDE (\ref{eq:spde}) for $\nu=0.1$ and $f=0$ numerically using the exponential Euler
scheme \citep{jentzen2009overcoming} for the finite-dimensional projection (\ref{eq:finite_galerkin_SDE}).
For (\ref{eq:finite_galerkin_SDE}), we use a Fourier truncation with
$L=64$ i.e.~$-64\leq k_{1},k_{2}\leq 64$. For $\pi_{0}$ we use
$\beta=0.5$, $\alpha=3$ and $\mu=v_{0}^{\dagger}$, with $v_{0}^{\dagger}$
being a random sample from $\mathcal{N}(0,A^{-\alpha})$ that is also used as the true signal to generate the observations.
To determine $Q$ we use $\sigma_{k}=\sqrt{2\delta\nu}|k|^{-3}$
with $\delta=1$. For the observation equation in
(\ref{eq:obs_eqn}) we use $\Sigma=0.8I$ and for the observer in
(\ref{eq:observer}) we place the observers' locations $x_{l}$
on a uniform square grid with equal spacing and set $r$ to be small
(smaller than $2\pi/L$). Thus, we can make the likelihood
more informative by decreasing the observation noise or by increasing
the grid size. As the  information in  the likelihood increases, one
expects a larger number of tempering steps (and slower total execution
times). When no tempering is used this will lead to a much lower value
for the ESS.

We  present results from two types of experiments with simulated
observations. In the first case we will look at a batch of $n=5$
observations from a dense grid ($16\times16$). We use this short
run to illustrate the efficiency and performance of the methodology.
The length of the data-set allows using multiple independent runs
for the same observations. In the second experiment we   use a
large number of observations ($n=100$) obtained from a $8\times8$
grid using both Gaussian and Student-t distributed additive noise.
We   show that the method performs well for the longer time and 
performance is similar for both Gaussian and non-Gaussian
observations. 

We begin with the case of $n=5$ and  dense observation grid
($16\times16$). In Table \ref{tab:imp-sam-efficiency-temper} we
present results for $N=100$ and $\delta t_{n}=0.4$
comparing a naive bootstrap PF, a PF that uses the informed proposal (\ref{eq:proposal})
for IS but without tempering (both based on Algorithm \ref{alg:smc-1}),
a PF that uses tempering when sampling from the stochastic NSE dynamics
in (\ref{eq:spde}), and a PF that uses both tempering and (\ref{eq:proposal})
for IS. We show the number of tempering steps per batch of observations,
the $ESS$ at each observation time $t_{p}$, and $L^{2}$-errors between the true
signal vorticity $w^{\dagger}$ and the estimated posterior mean $\hat{w}$ at each
epoch, i.e.
\begin{equation*}
\int_{\Omega}\left\Vert \hat{w}(x,t_{i})-w^{\dagger}(x,t_{i})\right\Vert ^{2}dx.
\end{equation*}
{For the $L^{2}$-errors we also include in Table \ref{tab:imp-sam-efficiency-temper} results from a standard Ensemble Kalman Filter (EnKF) \citep{evensen2009data}. 
It should be noted that the EnKF is computationally cheaper and usually it is used with lower values for $N$ than here. 
We include it not for the sake of a direct comparison, but to provide a benchmark for 
performance. %For instance, it shows the improvement when IS is used compared to bootstrap PF. 
}

\begin{table}[t]
\centering
{\footnotesize
\begin{tabular}{|c|c|c|c|c|c||c|c|c|c|c|}
\hline 
 & \multicolumn{5}{c||}{{Tempering steps}} & \multicolumn{5}{c|}{{$ESS$ }}\tabularnewline
\hline 
{$n=$} & {$1$} & {$2$} & {$3$} & {$4$} & {$5$} & {$1$} & {$2$} & {$3$} & {$4$} & {$5$}\tabularnewline
\hline 
\hline 
{IS-PF-T} & {5.6} & {4.7} & {4.4} & {4} & {4.3} & {64.87 } & {73.88} & {63.02} & {57.01} & {53.03}\tabularnewline
\hline 
{PF-T} & {10.1} & {7.7} & {7.4} & {7.6} & {8.1} & {77.73} & {70.62} & {68.50} & {75.88} & {82.02}\tabularnewline
\hline 
{IS-PF } & \multicolumn{5}{c||}{{n/a}} & {1.16} & {1.92} & {1.56} & {2.11} & {1.90}\tabularnewline
\hline 
{PF} & \multicolumn{5}{c||}{{n/a}} & {1.00} & {1.00} & {1.01} & {1.13} & {1.06}\tabularnewline
\hline 
\end{tabular}{}%
{\medskip{}
}{\footnotesize \par}
\begin{tabular}{|c|c|c|c|c|c|}
\hline 
 & \multicolumn{5}{c|}{{$L^{2}$-errors}}\tabularnewline
\hline 
{$n=$} & {$1$} & {$2$} & {$3$} & {$4$} & {$5$}\tabularnewline
\hline 
\hline 
{IS-PF-T} & {$0.19$ $(0.0012)$} & {$0.26$ $(0.0002)$} & {$0.21$ $(0.0003)$} & {$0.16$ $(0.0001)$} & {$0.27$ $(0.0005)$}\tabularnewline
\hline 
{PF-T} & {$0.43$ $(0.0110)$} & {$0.32$ $(0.0029)$} & {$0.25$ $(0.0054)$} & {$0.23$ $(0.0023)$} & {$0.38$ $(0.0137)$}\tabularnewline
\hline 
{IS-PF } & {$0.31$ $(0.0033)$} & {$0.45$ $(0.0166)$} & {$0.42$ $(0.0062)$} & {$0.33$ $(0.0062)$} & {$0.46$ $(0.0023)$}\tabularnewline
\hline 
{PF} & {$0.85$ $(0.0185)$} & {$1.13$ $(0.1493)$} & {$0.86$ $(0.0810)$} & {$0.96$ $(0.0260)$} & {$1.15$ $(0.0467)$}\tabularnewline
\hline 
{EnKF} & {$0.66$ $(0.1151)$} & {$0.60$ $(0.0108)$} & {$0.65$ $(0.0194)$} & {$0.63$ $(0.0245)$} & {$0.74$ $(0.0138)$}\tabularnewline
\hline 
\end{tabular}{\footnotesize \par}}
\caption{Average results for number of tempering steps, $ESS$ and
$L^{2}$-errors (with standard deviations in parenthesis) obtained
from $10$ independent runs of each algorithm. IS-PF-T denotes using the guided proposal with tempering, PF-T is bootstrap with tempering. 
The other two methods follow similarly and do not use tempering and MCMC steps.
In all cases we use $N=100$, $\delta t_{n}=0.4$. For the PF-T  we use $m=20$ MCMC steps (in Algorithm \ref{alg:mcmc}) with $\rho=0.9$ and for IS-PF-T we use $m=10$ and $\rho=0.5$. 
For $n=1$ we also use a pCN proposal for $V(0)$ that is invariant to $\pi_0$ with the step sizes being $\rho_0=0.98$ for PF-T and $\rho_0=0.9$ for IS-PF-T.}
\label{tab:imp-sam-efficiency-temper}
\end{table}

\begin{figure}[t]
\centering
\includegraphics[width=1.1\textwidth]
{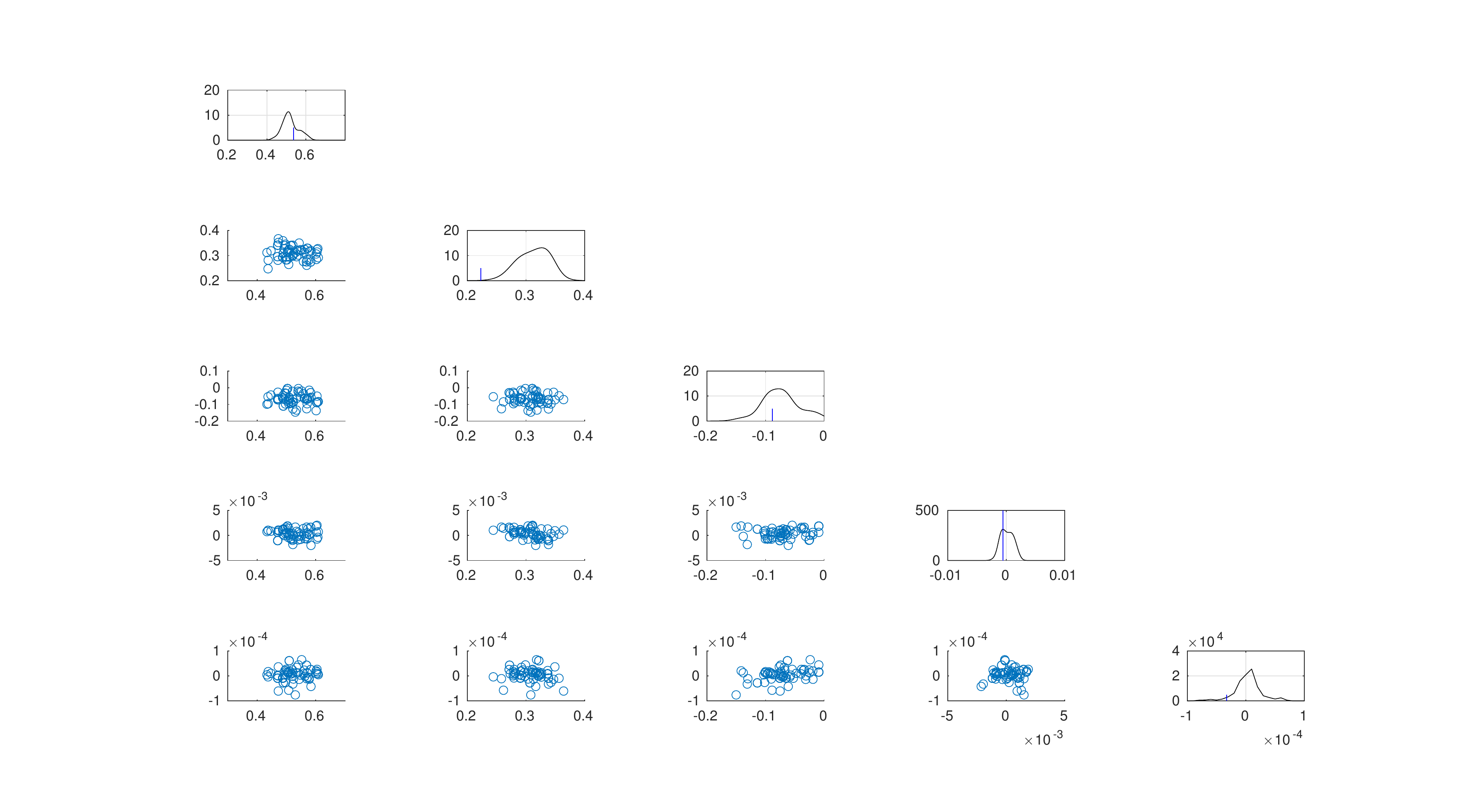}
\includegraphics[width=1.1\textwidth]{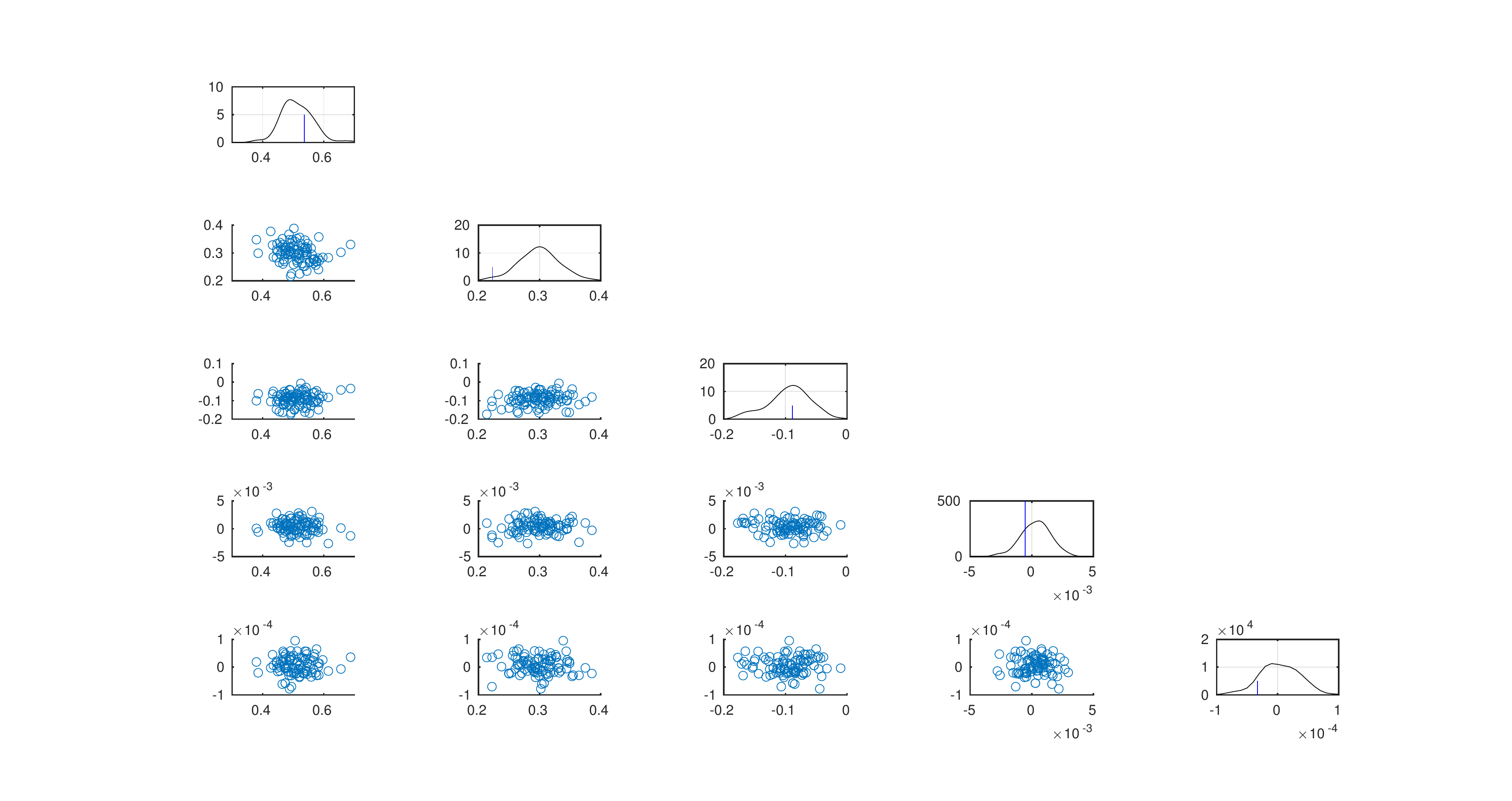}
\caption{PDF and scatter plots for $\text{Real}(u_{k})|\mathcal{Y}_{n}$ at	
$n=5$ for $k=(1,0),(1,1),(1,-1),(2,5),(9,9)$. Top is boostrap (sampling with (\ref{eq:spde})) and
bottom is IS (with (\ref{eq:proposal})) and both use tempering. Vertical
lines in PDF plots are true signal values used to generate the observations.}
\label{fig:scatter_IS_temper_n5} 
\end{figure}

When tempering is used we present in Figure \ref{fig:scatter_IS_temper_n5}
selected typical { estimated} PDFs and scatter plots for a few chosen frequencies $k$-s. In the scatter plots
the advantage
of using (\ref{eq:proposal}) ({in the bottom plot of Figure \ref{fig:scatter_IS_temper_n5}}) results in higher
dispersion of the particles relative to sampling from (\ref{eq:spde})
{(top plot). This is also apparent in the tails of the estimated PDFs.}
In Table \ref{tab:imp-sam-efficiency-temper}
it is evident that when using tempering, IS resulted in about half of the
tempering steps than when sampling from (\ref{eq:spde}). In both
cases, the tuning of the MCMC steps lead to the same acceptance ratio
(around $0.2$ at the final tempering step). We use $m=20$ MCMC iterations
per tempering for $n=1$. For $n>1$, plain tempering uses $m=20$ and IS with (\ref{eq:proposal}) uses $m=10$.
In addition, the IS-tempering case uses a larger step size (smaller $\rho$) for the MCMC (with $\rho=0.5$ rather than $0.9$).
This results in lower total computational cost and runtimes when IS
is used despite the added computations imposed by computing $g$ in 
(\ref{eq:proposal}). We also note that a lower number of tempering steps is
beneficial in addressing potential path degeneracy issues.
{
In Table \ref{tab:imp-sam-robust} we present results 
when $\delta t_{n}=0.16,1$ and $\Sigma=0.16I,0.4I$ to illustrate the robustness of Algorithm \ref{alg:smc} w.r.t 
spacing of observation times and signal to noise ratio. As expected more tempering steps are needed in the more informative observation case ($\Sigma=0.16I$),
but at the same time accurate observations result in lower $L^2$ errors. In addition, our method seems to perform comparatively better
when $\delta t_n=1$. This can be attributed to the guided proposal being given more time to evolve and guide the particles to better regions of the state space.}

\begin{table}[t]
{\footnotesize
\centering
\begin{tabular}{|c|c|c|c|c|c||c|c|c|c|c|}
\hline 
 & \multicolumn{5}{c||}{{Tempering steps }} & \multicolumn{5}{c|}{{$ESS$ }}\tabularnewline
\hline 
{$n=$} & {$1$} & {$2$} & {$3$} & {$4$} & {$5$} & {$1$} & {$2$} & {$3$} & {$4$} & {$5$}\tabularnewline
\hline 
\begin{tabular}{c}{$\delta t_{n}=0.4$},\\{$\Sigma=0.8I$} \end{tabular} & {5.6} & {4.7} & {4.4} & {4} & {4.3} & {64.87 } & {73.88} & {63.02} & {57.01} & {53.03}\tabularnewline
\hline 
\begin{tabular}{c}{$\delta t_{n}=0.16$},\\{$\Sigma=0.8I$}\end{tabular} & {5.9} & {4.3} & {4.0} & {4.0} & {4.1} & {54.16} & {53.79} & {40.29} & {90.06} & {54.53}\tabularnewline
\hline 
\begin{tabular}{c}{$\delta t_{n}=1$},\\{$\Sigma=0.8I$}\end{tabular} & {5.1} & {4.6} & {4.6} & {4.6} & {5.1} & {65.25} & {45.23} & {80.27} & {99.78} & {82.45}\tabularnewline
\hline 
\begin{tabular}{c}{$\delta t_{n}=0.4$},\\{$\Sigma=4I$}\end{tabular} & {4.6} & {3.7} & {3.2} & {3.1} & {3.9} & {39.97} & {88.61} & {48.99} & {49.21} & {80.08}\tabularnewline
\hline
\begin{tabular}{c}{$\delta t_{n}=0.4$},\\{$\Sigma=0.16I$}\end{tabular} & {7.2} & {6.0} & {5.9} & {5.7} & {6.1} & {65.27} & {74.57} & {51.94} & {54.57} & {50.26}\tabularnewline
\hline
\end{tabular}{}%
\medskip{}
\par
\begin{tabular}{|c|c|c|c|c|c|}
\hline 
 & \multicolumn{5}{c|}{{$L^{2}$-errors}}\tabularnewline
\hline 
{$n=$} & {$1$} & {$2$} & {$3$} & {$4$} & {$5$}\tabularnewline
\hline 
\begin{tabular}{l}{$\delta t_{n}=0.4$},\\{$\Sigma=0.8I$}\end{tabular}& {$0.19$ $(0.0012)$} & {$0.26$ $(0.0002)$} & {$0.21$ $(0.0003)$} & {$0.16$ $(0.0001)$} & {$0.27$ $(0.0005)$}\tabularnewline
\hline 
\begin{tabular}{l}{$\delta t_{n}=0.16$},\\{$\Sigma=0.8I$}\end{tabular}& {$0.24$ $(0.0004)$} & {$0.22$ $(0.0018)$} & {$0.29$ $(0.0013)$} & {$0.26$ $(0.0019)$} & {$0.24$ $(0.0015)$}\tabularnewline
\hline 
\begin{tabular}{l}{$\delta t_{n}=1$},\\{$\Sigma=0.8I$}\end{tabular}& {$0.27$ $(0.0004)$} & {$0.28$ $(0.0005)$} & {$0.24$ $(0.0004)$} & {$0.15$ $(0.0001)$} & {$0.17$ $(0.00004)$}\tabularnewline
\hline 
\begin{tabular}{l}{$\delta t_{n}=0.4$},\\{$\Sigma=4I$}\end{tabular}& {$0.31$ $(0.0034)$} & {$0.61$ $(0.0057)$} & {$0.50$ $(0.0027)$} & {$0.38$ $(0.0021)$} & {$0.56$ $(0.0009)$}\tabularnewline
\hline 
\begin{tabular}{l}{$\delta t_{n}=0.4$},\\{$\Sigma=0.16I$}\end{tabular}& {$0.12$ $(0.0003)$} & {$0.10$ $(0.00009)$} & {$0.08$ $(0.00006)$} & {$0.07$ $(0.00001)$} & {$0.12$ $(0.0001)$}\tabularnewline
\hline 
\end{tabular}{\footnotesize \par}
\caption{Average results of Algorithm \ref{alg:smc} (IS-PF-T) when varying $\Sigma$ and $\delta t_{n}$. Results are from $10$ independent runs and presented similarly to Table \ref{tab:imp-sam-efficiency-temper}. 
For the MCMC step sizes we use $\rho_0=0.9$ (for $\pi_0$) and $\rho=0.5,\:0.5,\:0.9,\:0.5,\:0.9$ for each case from top to bottom.}
\label{tab:imp-sam-robust}
}
\end{table}

\begin{figure}[t]
\centering\includegraphics[width=0.9\textwidth]{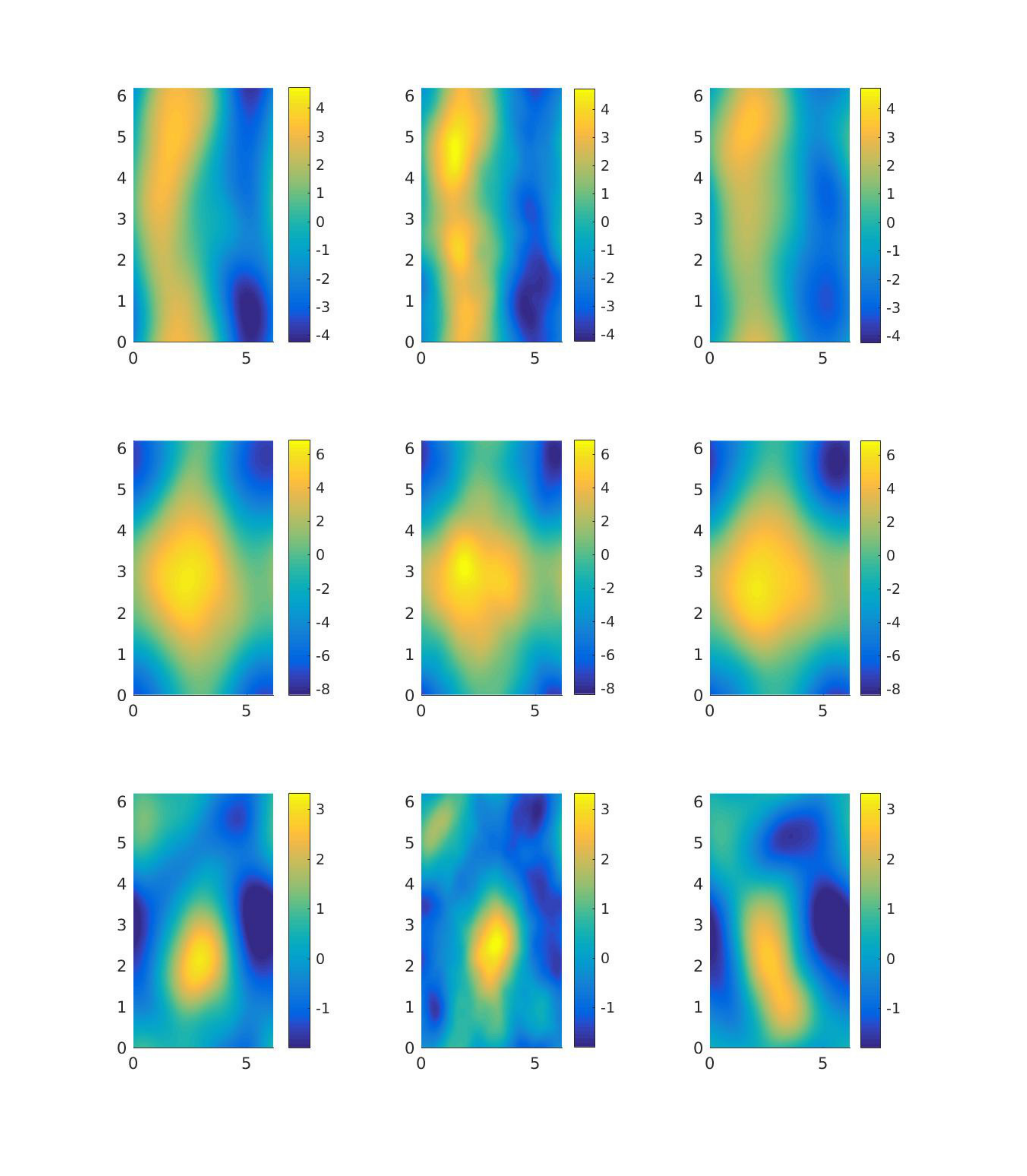}
\caption{Vorticity plots showing posterior mean of $p(\nabla\times V_{t}|\mathcal{Y}_{n})$ and true signal: top row $n=10$, middle $n=50$, bottom $n=100$. The left column
contains posterior means from Gaussian observation noise, the right one from Student-t
noise and in the middle is $w_{t}^{\dagger}$ (true signal vorticity). }
\label{fig:vort_plots} 
\end{figure}

\begin{figure}[t]
\centering
\includegraphics[width=0.85\textwidth]{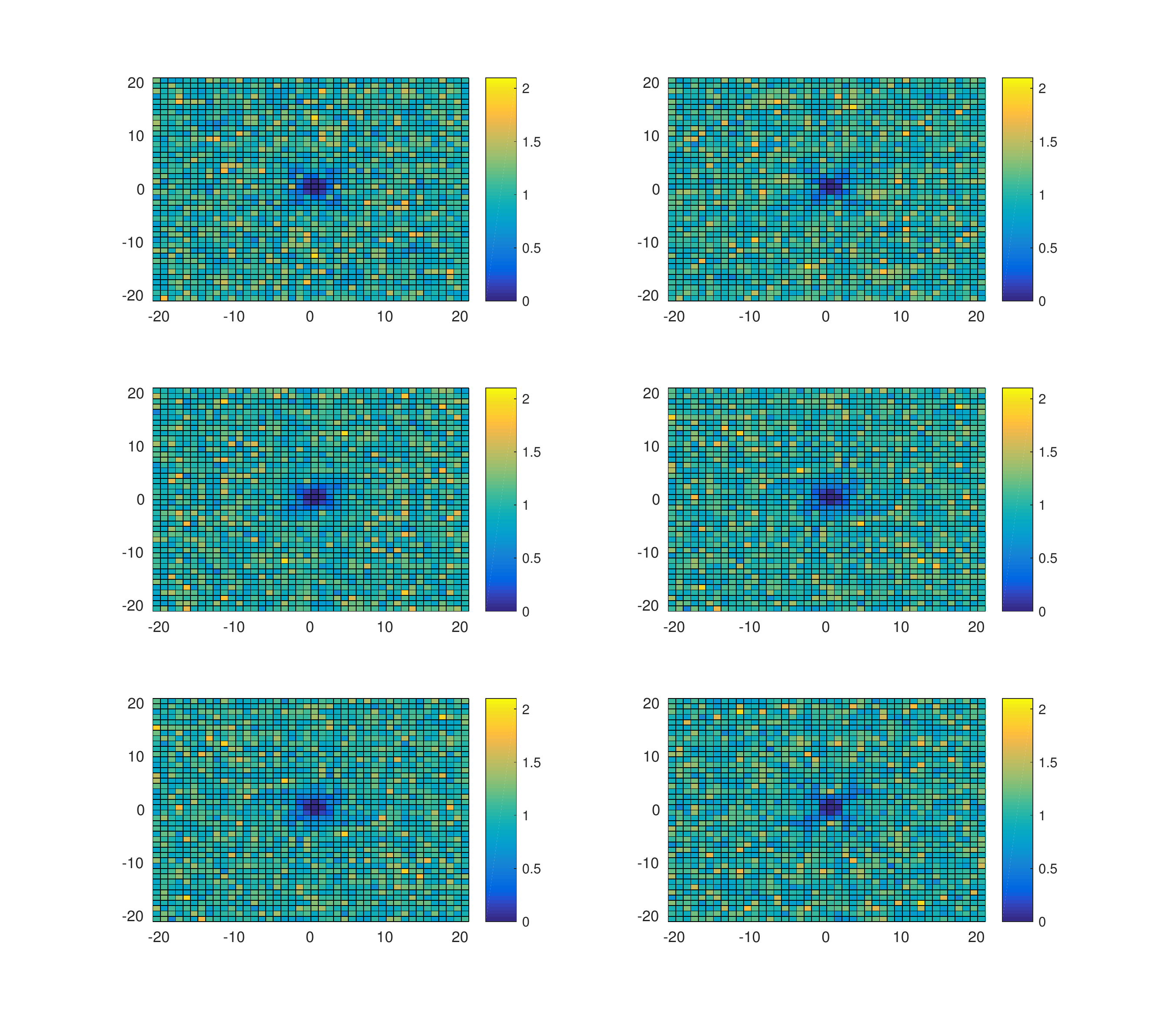}
\caption{Variance plots: top row $n=10$, middle $n=50$, botton $n=100$.We
present heat maps of the ratio of the posterior variance of $\pi_{t_n}$
over the variance for the law of the signal dynamics against $k$;
left part is for Gaussian noise and right for Student-t.}
\label{fig:vort_plots2} 
\end{figure}

\begin{figure}[t]
\centering\includegraphics[width=0.45\textwidth,height=0.28\textwidth]{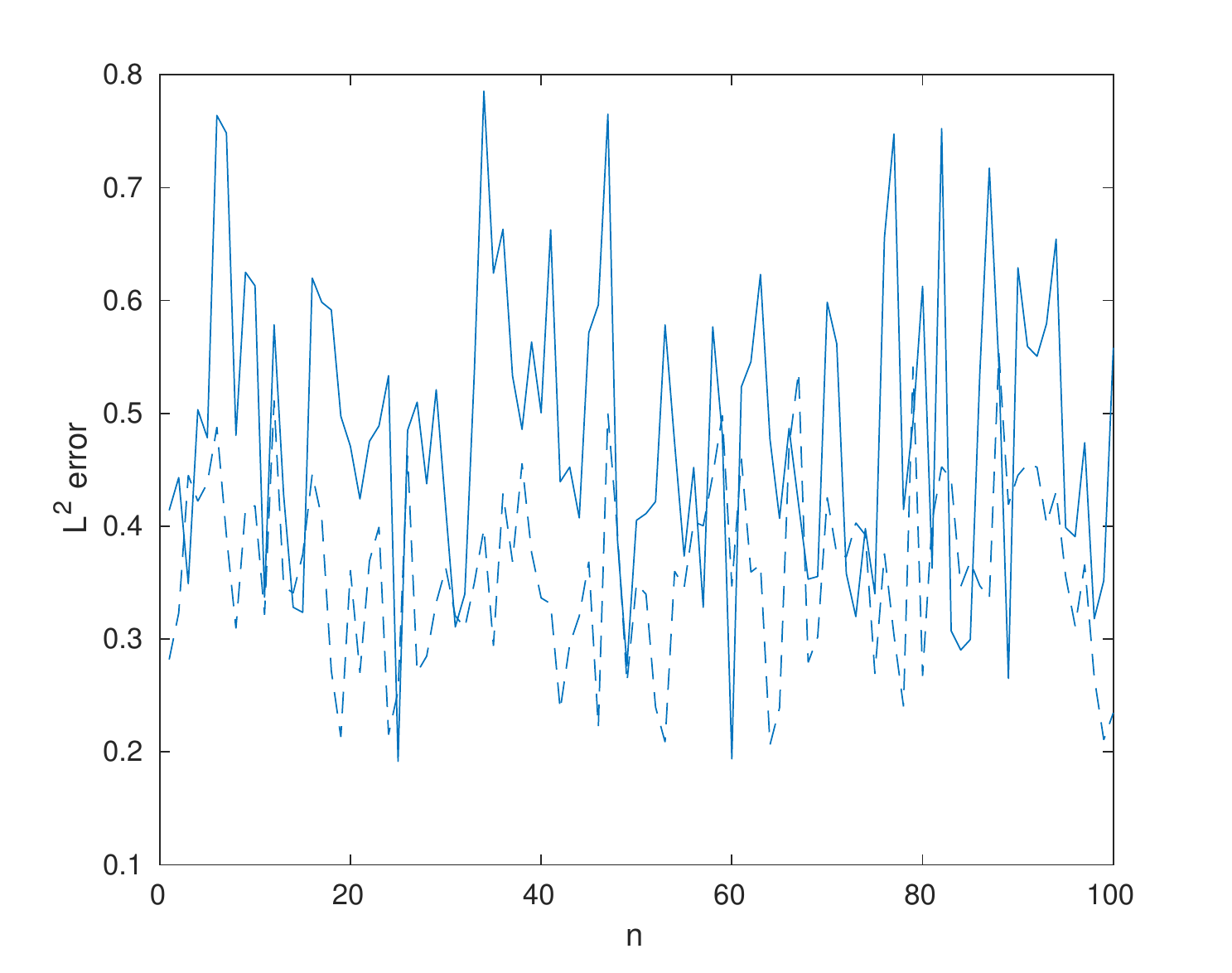}\includegraphics[width=0.45\textwidth,height=0.28\textwidth]{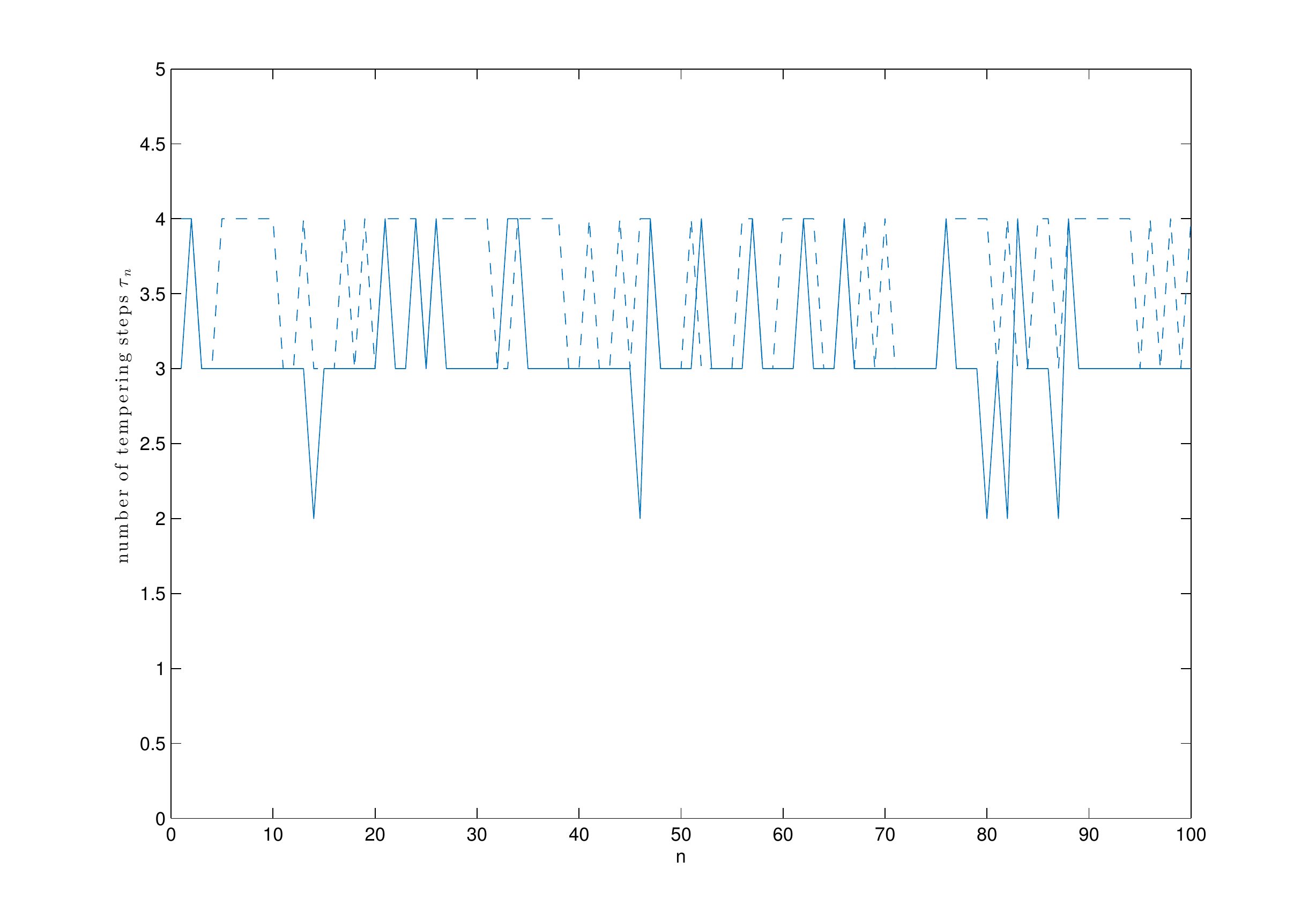}\\
 \includegraphics[width=0.45\textwidth,height=0.28\textwidth]{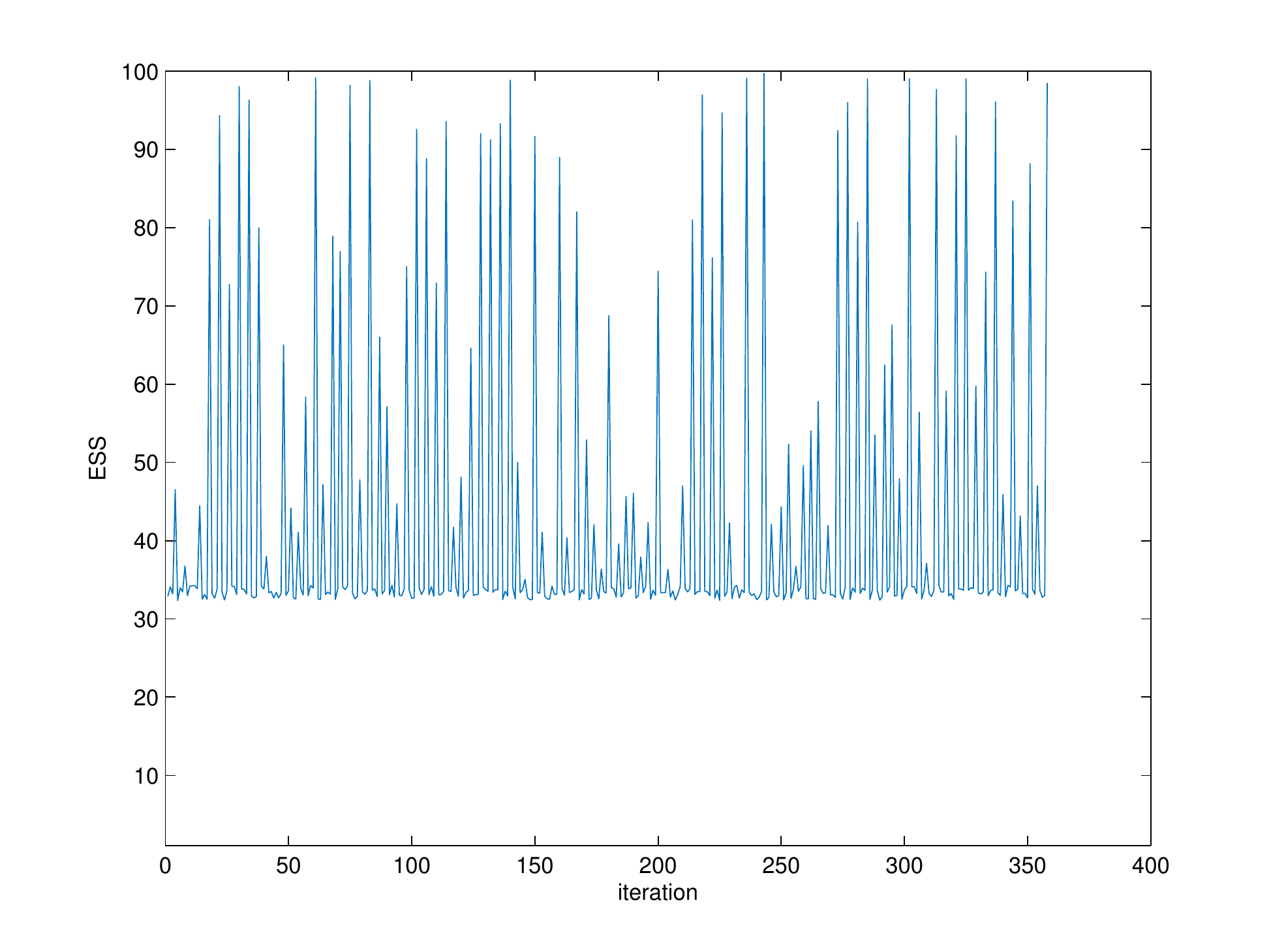}\includegraphics[width=0.45\textwidth,height=0.28\textwidth]{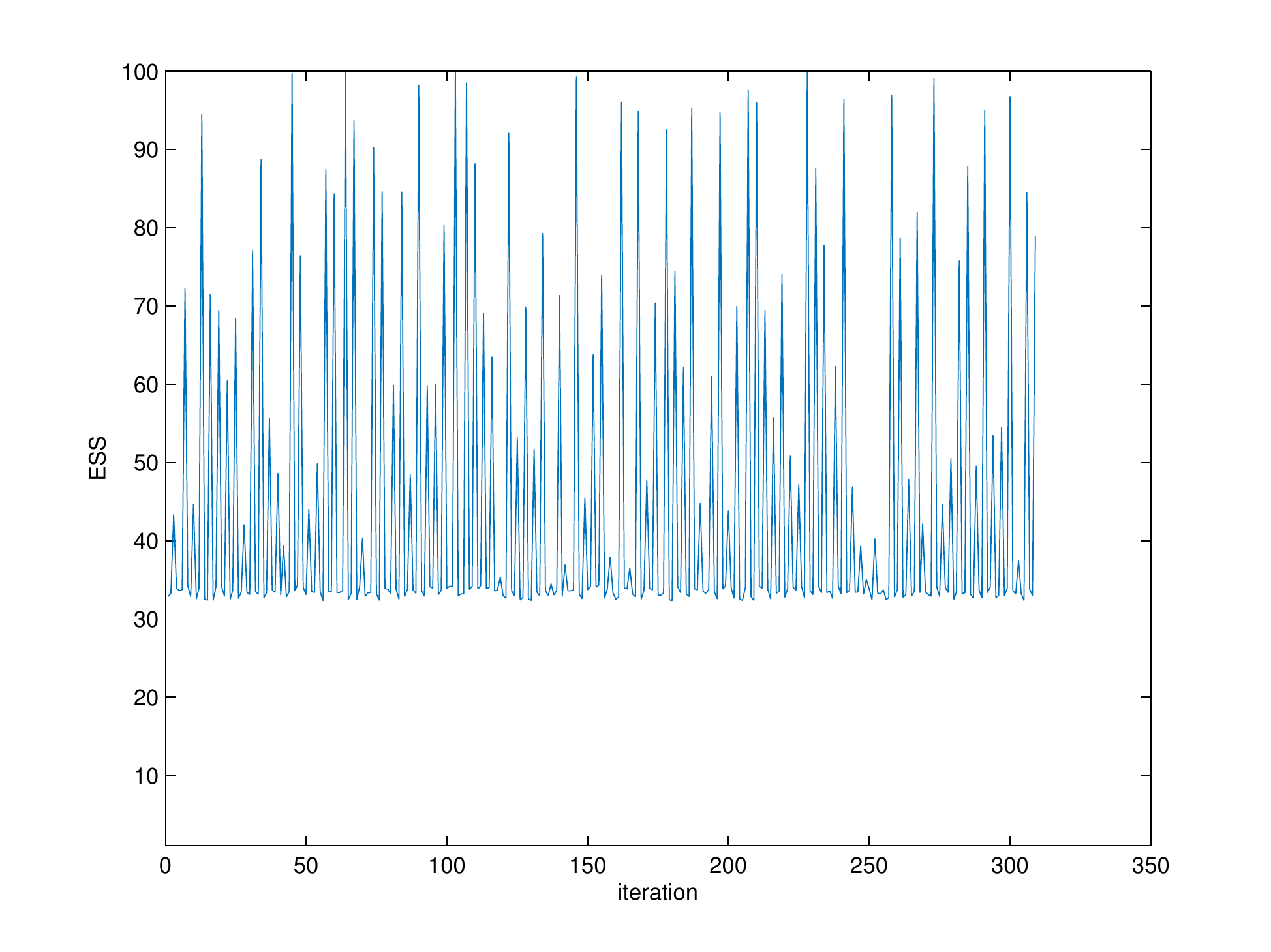}
\caption{Results for single run of PF with tempering and IS for $8\times8$
grid. Top panels are $L^{2}$-errors (left) and number of tempering steps
against $n$ (right). Dotted lines are for Gaussian observation noise and
solid for Student-t. In the bottom panels we present ESS against SMC
iteration for Gaussian (left) and Student-t (right) errors. Execution times
were $4.8018\times10^{5}$ and $4.17196\times10^{5}$ seconds respectively.}
\label{fig:long_vs_n_plots} 
\end{figure}

\begin{figure}[h]
\centering\includegraphics[width=0.9\textwidth,height=0.28\textwidth]{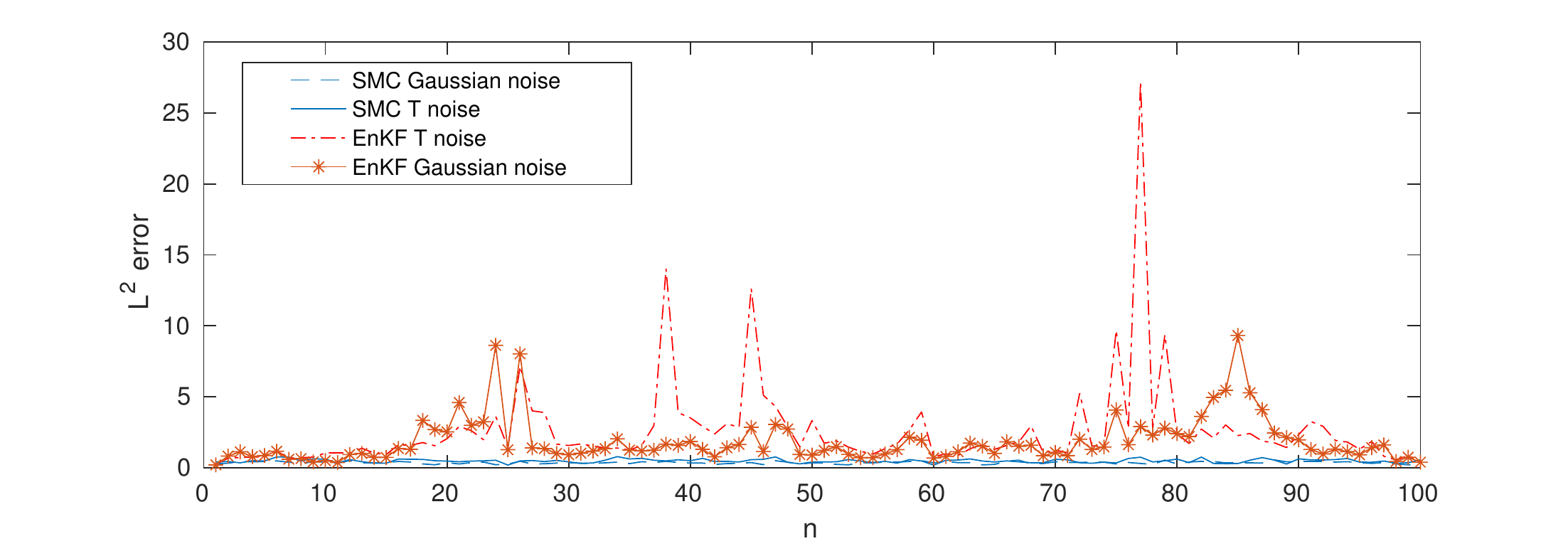}
\caption{$L^{2}$-error comparison for IS-PF with tempering and EnKF with Gaussian and
Student-t observation noise. PF errors are same as top left panel of Figure \ref{fig:long_vs_n_plots} and much lower then EnKF.}
\label{fig:L2_Error} 
\end{figure}

\begin{figure}[h]
\centering\includegraphics[width=0.33\textwidth]{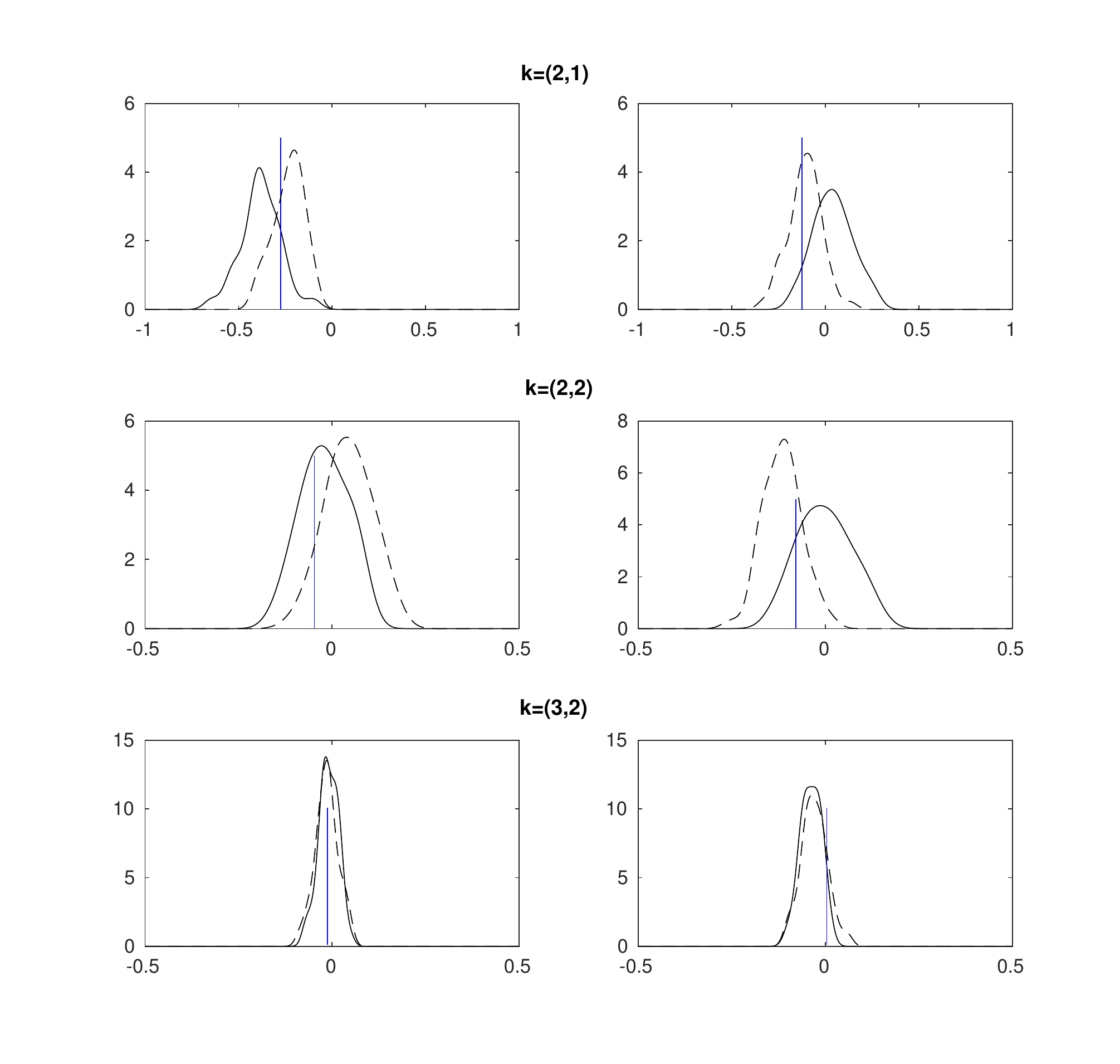}\vrule\includegraphics[width=0.33\textwidth]{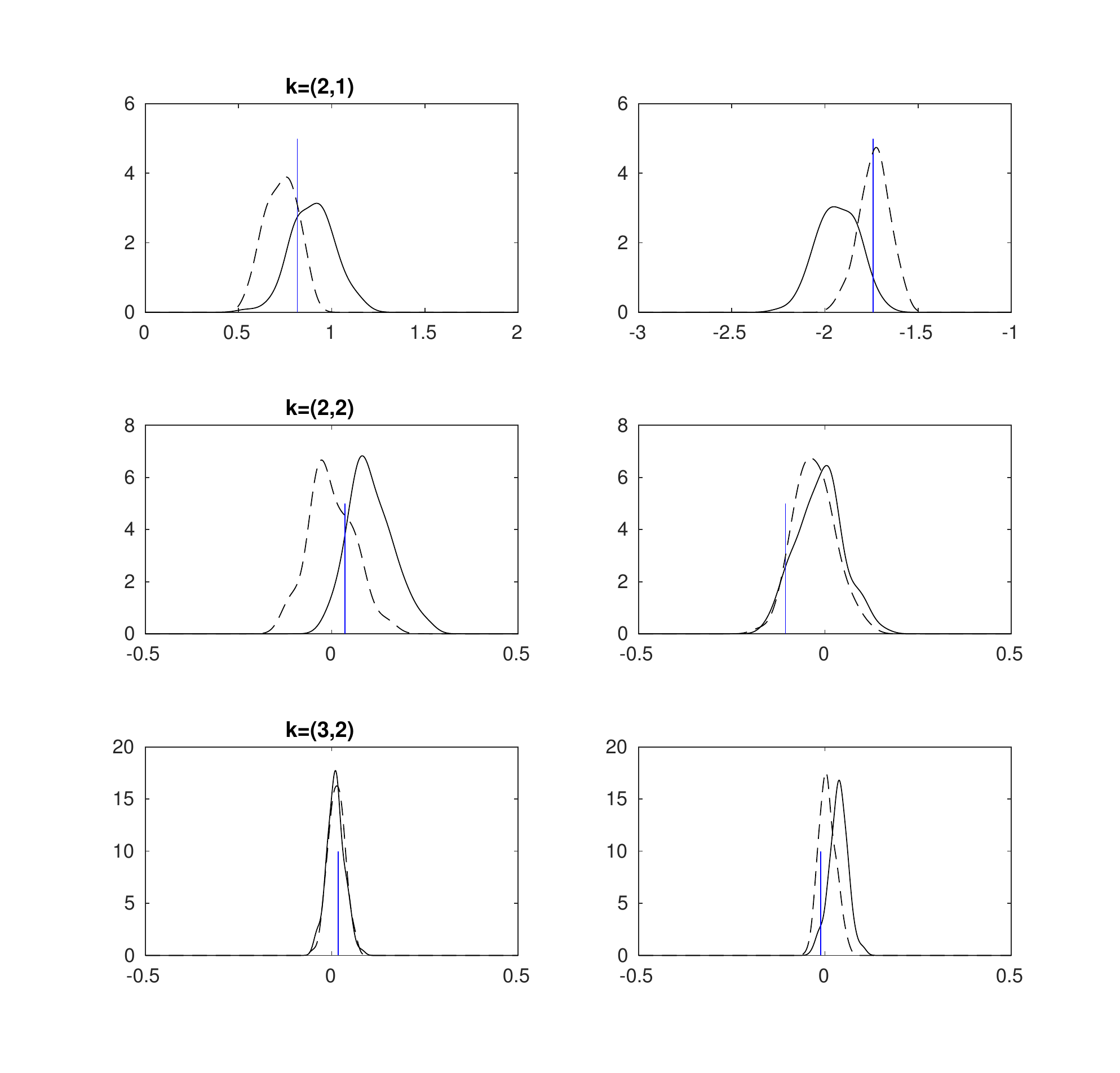}\vrule\includegraphics[width=0.33\textwidth]{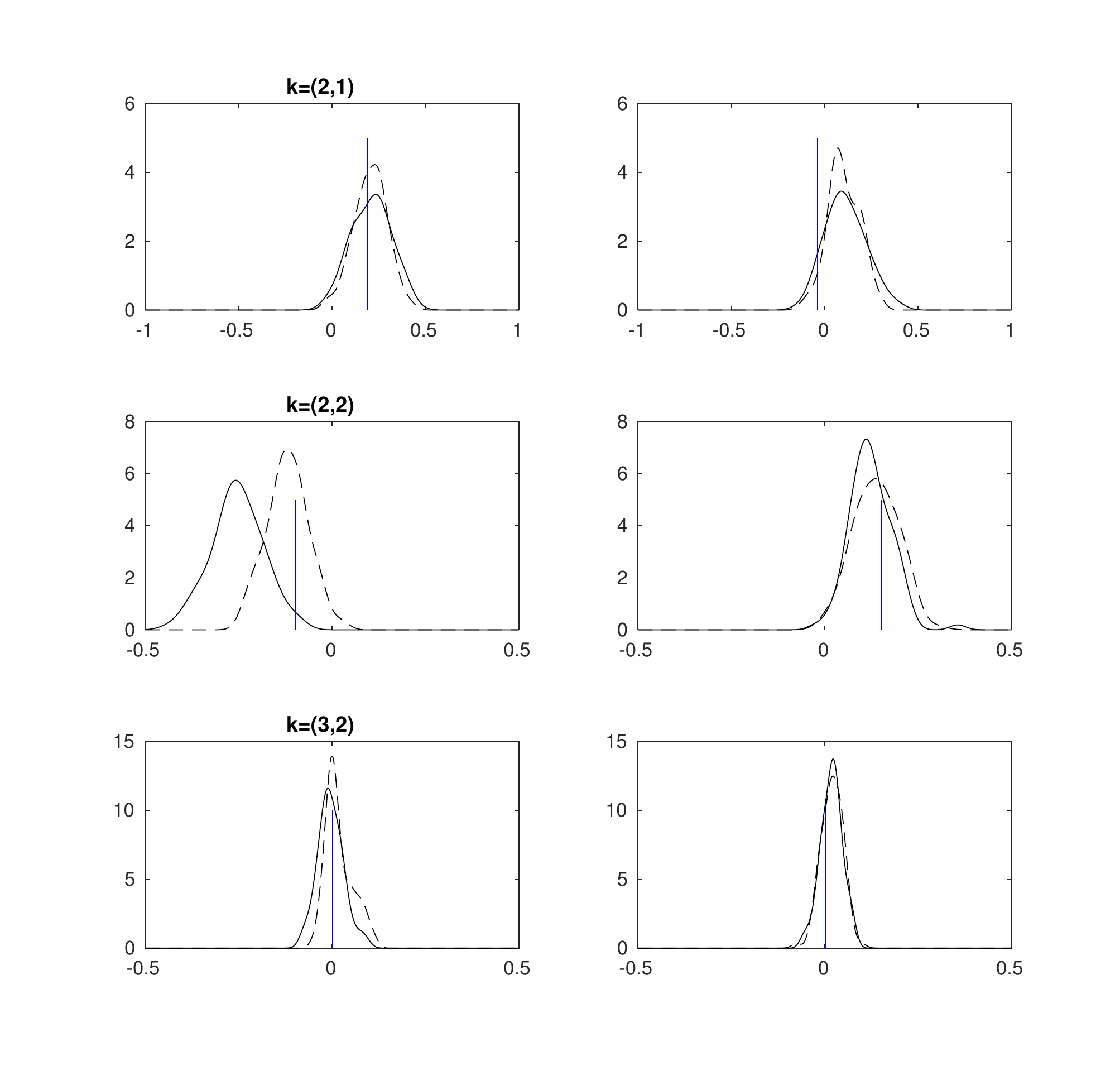}
\caption{PDFs of vorticity for $n=10,50,100$ from left to right. In each
panel for $n$, left side displays real part and right is imaginary;
top row is $k=(2,1)$, middle $k=(2,2)$ and bottom is $k=(3,2)$.
Dotted line is for Gaussian observation noise and solid for Student-t,
vertical lines are true signal values used to generate the observations.}
\label{fig:fixed_init_vort_mean-1} 
\end{figure}

We proceed with the second numerical experiment, where we use only
a single run of a PF with both tempering and IS for $N=100$ and $n=1,\ldots,100$. The dynamics for the state and true signal are as before, but for the observations we  
use a $8\times8$ equally spaced observation grid and look at
two different generated data-sets with different distributions for the noise $\Xi_{n}$: a zero mean
Gaussian and zero mean Student-t distribution with 4 degrees of freedom.
In both cases $\Sigma=0.8I$. {For each case, different PFs are implemented, each with the correctly specified likelihood.}
In Figure \ref{fig:vort_plots}
we plot the estimated vorticity posterior mean for $n=10,50,100$,
in each case together with the vorticity of the true signal. {The estimates seem accurate with small deviations between 
the posterior mean and the true signal. The latter is sensible given the coarseness of the grid and the moderate number of observations.} 
We also provide in Figure \ref{fig:vort_plots2} a plot of the ratio of the posterior variance of the vorticity
of $V_{t_{n}}$ over the unconditional variance when obeying the probability
law determined by the stochastic NSE dynamics in (\ref{eq:spde}). {The information gain appears as a reduction 
in the posterior variance for low $|k|$ relative to the prior, which is to be expected as the spatial grid cannot
be informative for higher wave-numbers.}
In Figure \ref{fig:long_vs_n_plots} we plot $ESS$, $L^{2}$-errors
as before and number of tempering steps per iteration. In both cases
the performance is fairly stable with time and the algorithm provides good posterior
mean estimates. {For completeness, in Figure \ref{fig:L2_Error} we include a comparison with the EnKF in terms of $L^2$ errors.
The PF with IS and tempering performs much better.
Finally, in Figure \ref{fig:fixed_init_vort_mean-1} we present some estimated
PDFs. These plots capture $\Pi_n$ for different $k$. Notice that the true parameter (displayed as a vertical line) 
lies in regions where the mass of the estimated posterior density is high 
and the posterior variance for the t-distributed case is higher for low $k$.}

\section{Discussion\label{sec:Conclusions}}
We have presented a particle filtering methodology that uses likelihood-informed IS proposals, tempering and MCMC moves for signals obeying
the stochastic NSE observed with additive noise. The approach is computationally
intensive and requires a significant number of particles $N$, but
we believe the cost is quite moderate relatively to the
dimensionality of the problem. The use of tempering and MCMC steps
is crucial for this high-dimensional application. The inclusion of
likelihood-informed proposals results in higher efficiency and $ESS$,
less tempering steps and higher step sizes for the MCMC steps - 
thus, overall, in lower computational cost. The IS proposals
were designed using a Gaussian noise assumption for the observations,
but we demonstrated numerically that they are still useful and efficient
for observation noise obeying a Student-t distribution with heavier
tails. In addition, as $\delta t_{n}$ increases using proposals as in
(\ref{eq:proposal}) will be more beneficial.

In the experiments presented in Section \ref{sec:Numerical-examples}
the effective dimensionality of the problem is determined by $\nu,{ \Sigma}$
and $\sigma_{k}.$ More challenging parameterizations than the ones
presented here could be dealt with by increasing $N$ or via a more advanced
numerical method for the solution of the SPDE.{ These can be addressed using 
the extensions discussed in Section \ref{sec:MCMCextensions}. Another} potentially useful extension is to
use different number
of particles for different ranges of $k$ following \citep{johansen2012exact}.
Furthermore we note that we did not make use of parallelization, but this is certainly possible
for many parts of Algorithm \ref{alg:smc} and can bring significant execution
speed-ups in applications. 

Future work could aim to extend this methodology by designing suitable IS proposals
for non-linear observation schemes or observations obtained from
Lagrangian drifters or floaters. {Finally, an interesting question is whether an error
analysis along the lines of \citep[Section 7.4]{del2004feynman} can be reproduced. The simulations presented 
here seem to indicate roughly constant errors with time, but a rigorous treatment would 
need to establish the stability properties 
of the filtering distribution w.r.t  the initialization.}

\section*{Acknowledgements}
FPLl was supported by EPSRC and the CDT in the Mathematics of Planet Earth under grant EP/L016613/1. AJ was supported by an AcRF tier 2 grant: R-155-000-161-112. AJ is
affiliated with the Risk Management Institute, the Center for Quantitative
Finance and the OR \& Analytics cluster at NUS. 
AB was supported by the Leverhulme Trust Prize.

\appendix

\section{More simulation results}

\begin{figure}
\centering
\includegraphics[width=1.1\textwidth]{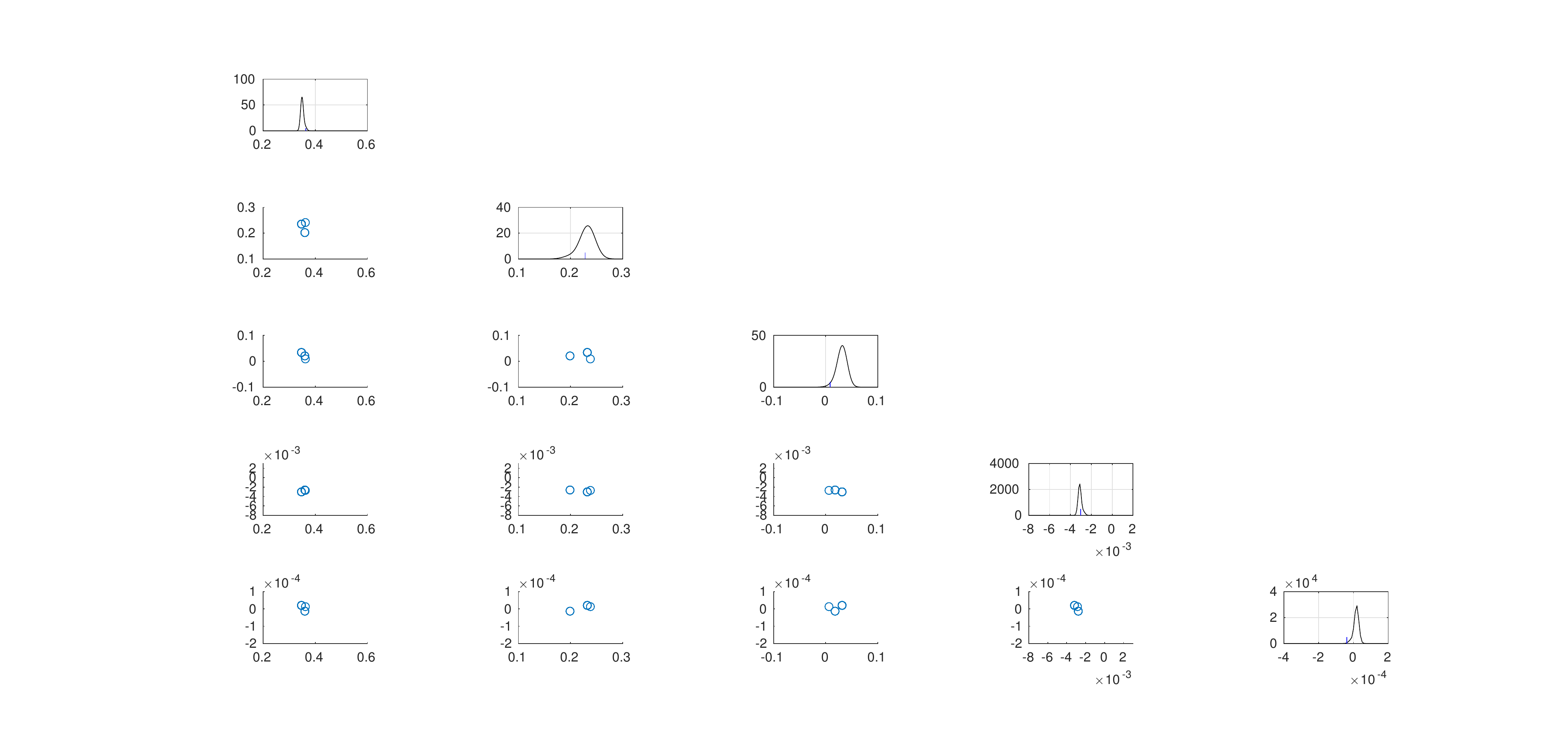}
\includegraphics[width=1.1\textwidth]{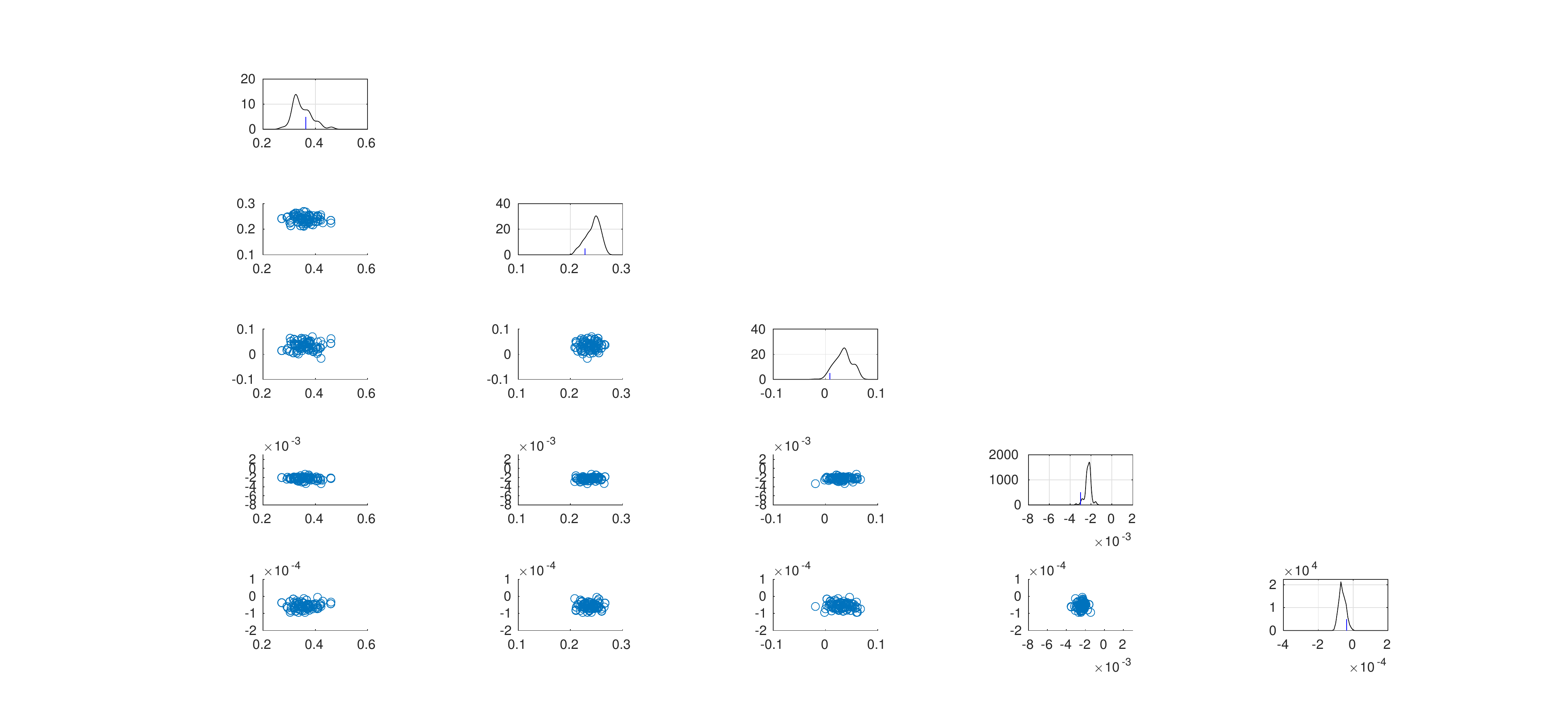}
\caption{Scatter plots at $n=5$ for perfect initialization $k=(1,0),(1,1),(1,-1),(2,5),(9,9)$.
Top is boostrap and bottom is IS with (\ref{eq:proposal}).}
\label{fig:fixed_init_scatter} 
\end{figure}

\begin{figure}
\centering\includegraphics[scale=0.29]{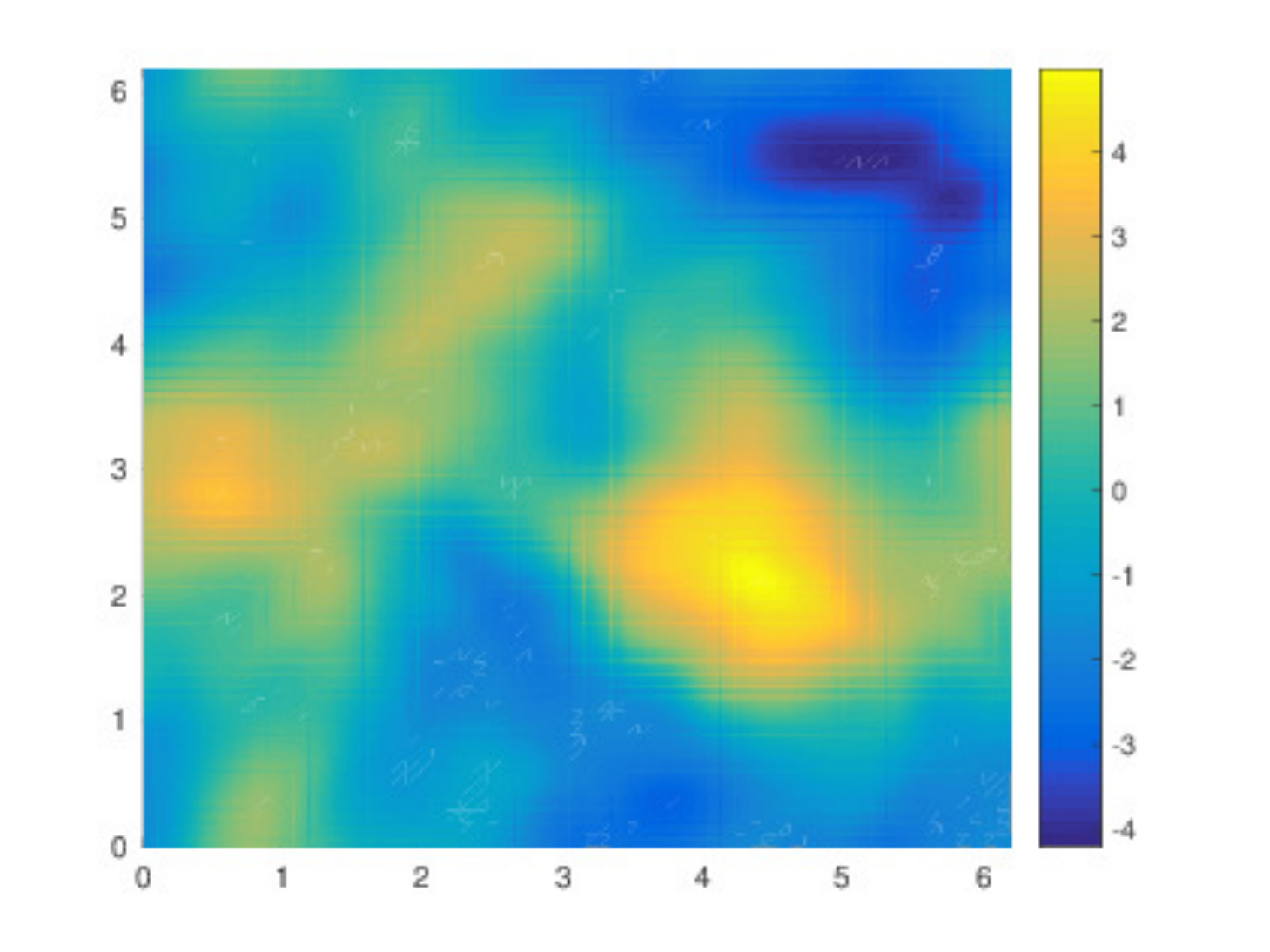}\includegraphics[scale=0.29]{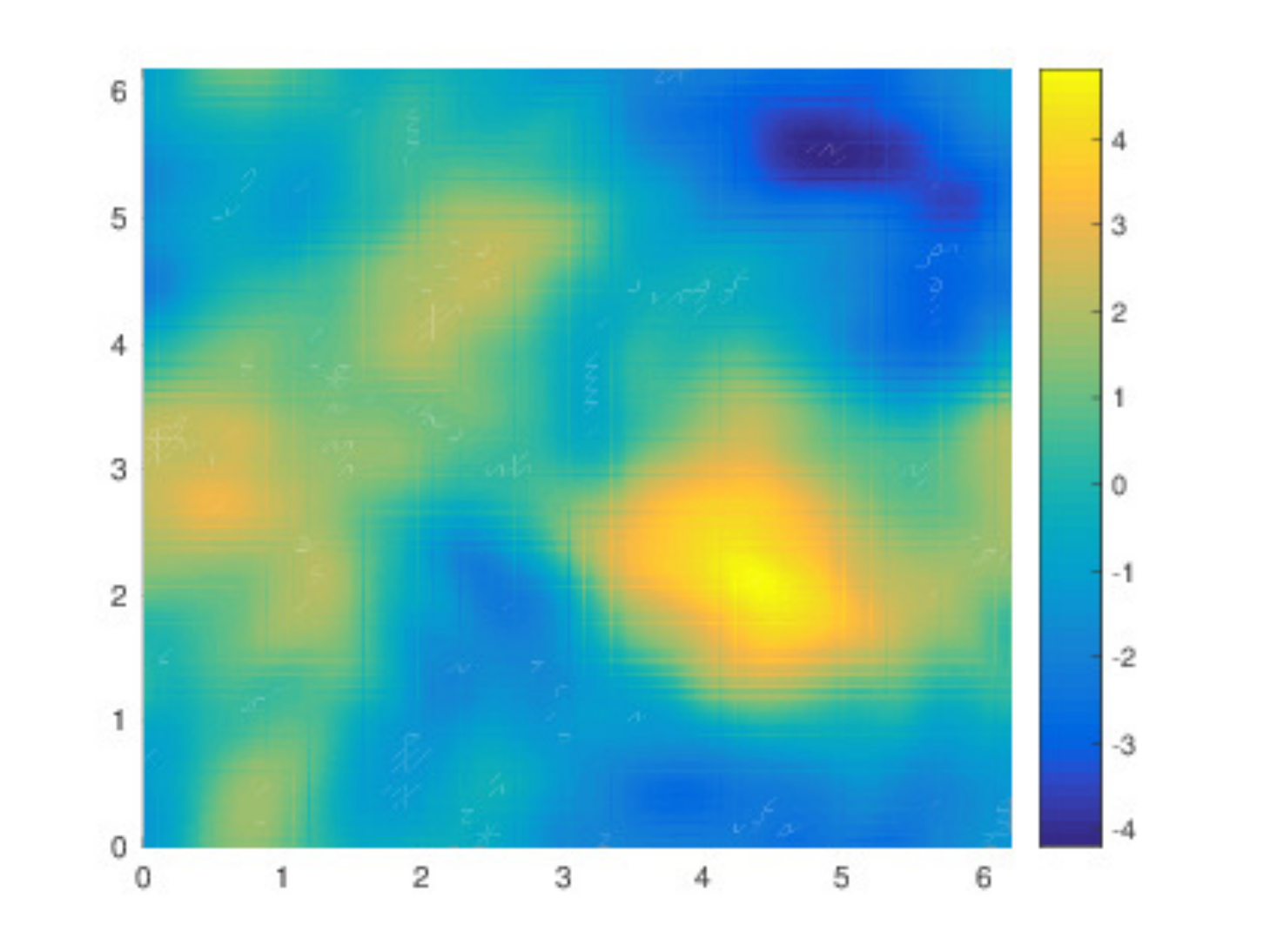}\includegraphics[scale=0.29]{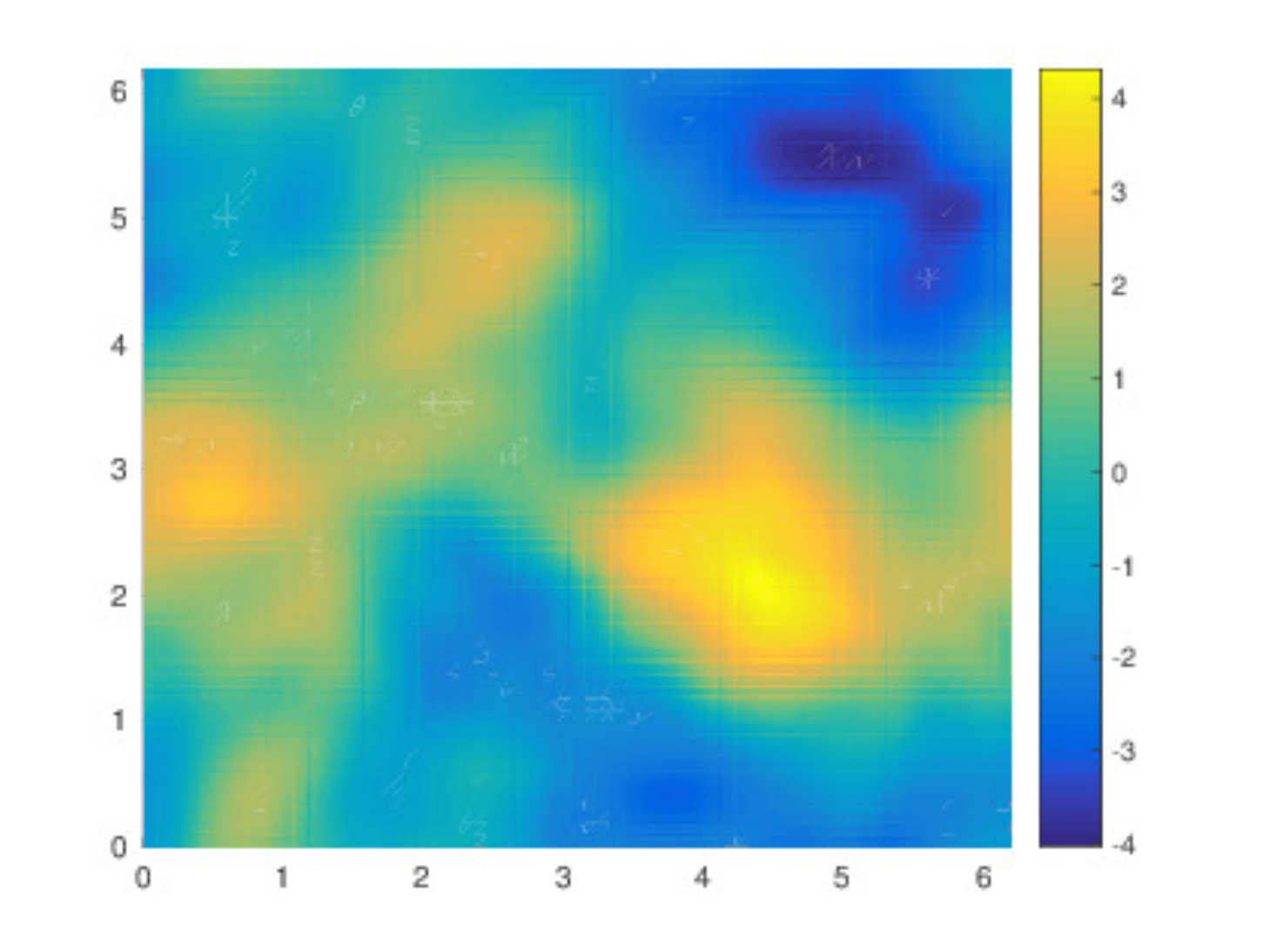}
\caption{Vorticity plots for $n=5$ and perfect initialization: left is posterior
mean from bootstrap PF, middle is real signal $v_{t}^{\dagger}$,
and right is posterior mean of PF of Algorithm \ref{alg:smc-1} and
IS with (\ref{eq:proposal}).}
\label{fig:fixed_init_vort_mean} 
\end{figure}

We present {some negative numerical results to illustrate that tempering is necessary}. 
We will consider a perfect initialization
for each particle with $v_{0}^{\dagger}$. Whilst this is an extremely favorable scenario
that is unrealistic in practice, it shows a clear benefit in using
IS and tempering. For $N=200$, $\nu=0.01$
and $\delta t_{n}=0.2$, we present some scatter plots in Figure \ref{fig:fixed_init_scatter}
for the experiment with $n=5$ seen earlier with a $16\times16$ block of observations. Notably, the estimated posterior means for the vorticity
seem to exhibit good performance; see Figure \ref{fig:fixed_init_vort_mean}.
Indicatively, the ESS here is $34$ for IS and $3$ for the bootstrap case.
Even in this extremely favorable scenario, the ESS is low and this strongly
motivates the use of tempering to improve the efficiency of the particle
methodology. In results not shown here, we also experimented the size of time increment $\delta t_{n}=t_{n}-t_{n-1}$ a naive
particle filter (Algorithm \ref{alg:smc-1}) can handle with perfect initialization. When the
likelihood-informed proposals in \eqref{eq:proposal} are used, the method produces
accurate {point estimates} for $\delta t_{n}$ up to $0.2-0.25$. This is in contrast
to when sampling from the dynamics, where the bootstrap version of
Algorithm \ref{alg:smc-1} can handle only up to $0.15$.

\bibliographystyle{siamplain}
\bibliography{navier}

\end{document}